\documentclass[10pt,journal,a4paper]{IEEEtran}
\usepackage{amssymb,amsmath}
\usepackage{cite}
\usepackage{graphicx, subfigure}
\usepackage{psfrag}
\usepackage{url}
\usepackage[latin1]{inputenc}
\usepackage[absolute,overlay]{textpos}
\usepackage{textcomp}
\usepackage{gensymb}
\usepackage{bm}

\usepackage{pgf}

\usepackage[nocomma]{optidef}

\usepackage{algorithm}
\usepackage{algorithmic}

\makeatletter
\def\thickhline{%
  \noalign{\ifnum0=`}\fi\hrule \@height \thickarrayrulewidth \futurelet
   \reserved@a\@xthickhline}
\def\@xthickhline{\ifx\reserved@a\thickhline
               \vskip\doublerulesep
               \vskip-\thickarrayrulewidth
             \fi
      \ifnum0=`{\fi}}
\makeatother

\newlength{\thickarrayrulewidth}
\usepackage{amsthm}

\begin{document}
    \title{Wireless Energy Transfer Beamforming Optimization for Intelligent Transmitting Surface}
\author{
	\IEEEauthorblockN{Osmel Mart\'{i}nez Rosabal, Onel Alcaraz L\'{o}pez, Victoria Dala Pegorara Souto, Richard Demo Souza, Samuel Montejo-S\'{a}nchez, Robert Schober, Hirley Alves}
        \thanks{This work is partially supported by UPRISING  (Grant Number 348515) and 6G Flagship (Grant Number 369116) funded by the Research Council of Finland; European Commission through the Horizon Europe/JU SNS projects Hexa-X-II (Grant no. 101095759) and AMBIENT-6G (Grant 101192113); CNPq (305021/2021-4), RNP/MCTI Brazil 6G (01245.020548/2021-07); FAPEMIG under projects No APQ-04523-23 and PPE-00124-23; XGM-AFCCT-2024-4-1-1, supported by xGMobile-EMBRAPII-Inatel Competence Center on 5G and 6G Networks, with financial resources from the PPI IoT/Manufacturing 4.0 program of MCTI grant number 052/2023, signed with EMBRAPII; and ANID FONDECYT Regular No. 1241977. The authors also wish to acknowledge CSC - IT Center for Science, Finland, for computational resources.}
        \thanks{Osmel Mart\'{i}nez Rosabal, Onel Alcaraz L\'{o}pez, and Hirley Alves are with the Centre for Wireless Communications (CWC), University of Oulu, Finland. Email: firstname.lastname@oulu.fi}
        \thanks{Victoria Dala Pegorara Souto is with the National Institute of Telecommunication, Santa Rita do Sapuca\'{i}, MG, Brazil. (victoria.souto@inatel.br).}	 
        \thanks{Richard Demo Souza is with the Department of Electrical and Electronics Engineering, Federal University of Santa Catarina, Florianopolis, SC, Brazil, (e-mail: richard.demo@ufsc.br).}
        \thanks{Samuel Montejo-S\'{a}nchez is with Instituto Universitario de Investigaci\'{o}n y Desarrollo Tecnol\'{o}gico, Universidad Tecnol\'{o}gica Metropolitana, Santiago, Chile. (smontejo@utem.cl).}
        \thanks{Robert Schober is with the Friedrich-Alexander University of Erlangen-Nuremberg, Germany. (robert.schober@fau.de)}
}   
\maketitle

\begin{abstract}
Radio frequency (RF) wireless energy transfer (WET) is a promising technology for powering the growing ecosystem of Internet of Things (IoT) using power beacons (PBs). Recent research focuses on efficient PB architectures that can support numerous antennas. In this context, PBs equipped with intelligent surfaces present a promising approach, enabling physically large, reconfigurable arrays. Motivated by these advantages, this work aims to minimize the power consumption of a PB equipped with a passive intelligent transmitting surface (ITS) and a collocated digital beamforming-based feeder to charge multiple single-antenna devices. To model the PB's power consumption accurately, we consider power amplifiers nonlinearities, ITS control power, and feeder-to-ITS air interface losses. The resulting optimization problem is highly nonlinear and nonconvex due to the high-power amplifier (HPA), the received power constraints at the devices, and the unit-modulus constraint imposed by the phase shifter configuration of the ITS. To tackle this issue, we apply successive convex approximation (SCA) to iteratively solve convex subproblems that jointly optimize the digital precoder and phase configuration. Given SCA's sensitivity to initialization, we propose an algorithm that ensures initialization feasibility while balancing convergence speed and solution quality. We compare the proposed ITS-equipped PB's power consumption against benchmark architectures featuring digital and hybrid analog-digital beamforming. Results demonstrate that the proposed architecture efficiently scales with the number of RF chains and ITS elements. We also show that nonuniform ITS power distribution influences beamforming and can shift a device between near- and far-field regions, even with a constant aperture.
\end{abstract}
\begin{IEEEkeywords}
Internet of Things, wireless energy transfer, nonlinear power amplifier, hybrid analog-digital beamforming, intelligent transmitting surface.
\end{IEEEkeywords}
\section{Introduction}\label{sec:intro}
Radio frequency (RF) wireless energy transfer (WET) has potential to provide a reliable and cost-effective energy supply to support the accelerated growth of Internet of Things (IoT) deployments \cite{Rosabal.2023,Lopez.2023}. WET relies on power beacons (PBs) which wirelessly recharge the IoT devices by RF waves. It provides several advantages such as \cite{Rosabal.2023} enabling new use cases and reducing e-waste by eliminating the need for battery replacements/recharging.

Notably, coverage remains challenging for WET-enabled networks due to their low end-to-end conversion efficiency and electromagnetic field exposure regulations, such as the International Commission on Non-Ionizing Radiation Protection guidelines \cite{Ziegelberger.2020}. Several techniques have been proposed to address these limitations, including energy beamforming, nomadic WET via flying/moving PBs, fluid antenna-enabled PBs, and the deployment of reconfigurable intelligent surfaces (RIS) \cite{Rosabal.2023}. Among these, energy beamforming plays a pivotal role in extending WET coverage by focusing energy in specific locations or angular directions through multi-antenna transmissions. However, fully exploiting the benefits of beamforming requires balancing complexity, power consumption, and deployment costs, which are key for the next generation of sustainable WET systems \cite{Rosabal.2023}.

Sustainability, beyond efficiency alone, has become a primary driver for next-generation WET-enabled networks, especially regarding efforts to reduce CO2 emissions. The next generation of sustainable WET systems envisions PB deployments relying solely on renewable energy, whose spatiotemporal variations may limit the PBs energy budget. Therefore, beamforming architectures with lower energy consumption enable more flexible deployments via nomadic WET and scale the network deployment to enhance coverage, especially in remote off-grid locations. A vision for aligning WET with current sustainable development goals is presented in \cite{Rosabal.2023}, while \cite{Lopez.2023} emphasizes key performance and value indicators needed to achieve this vision. 

\subsection{Cost-effective energy beamforming}\label{subsec:MIMO}
Efforts towards achieving cost-effective beamforming include i) the use of low-resolution analog-to-digital converters, whose power consumption scales with the number of quantization bits \cite{Khalili.2022}; ii) antenna selection architectures, in which the analog phase shifters are replaced by switches \cite{Elbir.2023}; and iii) algorithm implementations at the hardware level to reduce area and power consumption of the integrated circuits \cite{Mirfarshbafan.2021}. In addition, novel analog front-end architectures that exploit mutual coupling or passive feeding networks have emerged as promising low-complexity alternatives \cite{Sedaghat.2016}. These approaches reshape the radiation pattern by adjusting tunable passive loads rather than relying on dedicated RF chains, significantly reducing hardware complexity and power consumption.

In the realm of hybrid beamforming, recent research has focused on developing novel architectures that can efficiently handle numerous antennas with a limited number of RF chains \cite{Zhang.2020,Lopez.2023,Samaiyar.2023,Jamali.2021, Bereyhi.2020}. For instance, lens antenna arrays achieve angle-dependent energy beamforming by employing a set of transmit antennas located in the focal region of an electromagnetic lens \cite{Zhang.2020}. In principle, the lens focuses electromagnetic signals arriving from multiple directions onto distinct ports, effectively transforming the channel into a sparse beamspace representation. This approach enables the activation of fewer RF chains without compromising performance \cite{Zhang.2020}. Unlike traditional beamforming architectures, lens antenna arrays do not rely on lossy and expensive phase shifters, thus offering substantial hardware and power savings compared to phase shifters-based implementations \cite{Zhang.2020}. On the other hand, dynamic metasurface antennas are composed of multiple waveguides embedding a large set of radiating metamaterial elements whose frequency response can be externally and individually adjusted by varying the local electrical properties \cite{Lopez.2023}. A single RF chain feeds each microstrip, but the input signal is radiated by all the elements located on the microstrip.

RIS-equipped transmitters, in which the surface is embedded into the transmitter architecture, have also gained attention \cite{Samaiyar.2023,Jamali.2021,Bereyhi.2020}. In this architecture, a collocated digital feeder illuminates the RIS. Then, the impinging signals are reflected or refracted from the surface in the desired spatial directions determined by the phase shift configuration of its passive reflecting elements. The air interface between the feeder and the RIS replaces the otherwise expensive/complex analog RF networks utilized in equivalent hybrid analog-digital beamforming implementations. Unlike scenarios where the RIS is deployed far from the transmitter, e.g., \cite{Wu.2025}, resulting in a fading channel that requires estimation, RIS-equipped transmitters benefit from a fixed channel that can be engineered during manufacturing. For further details concerning RIS-equipped transmitters, \cite{Abdelrahman.2017} and \cite{Mirmozafari.2021} offer an in-depth theoretical analysis, while \cite{Wang.2022} showcases the beamforming capabilities of an electronically reconfigurable proof-of-concept transmitter. Unlike lens antenna arrays, the reconfigurability of RIS enables the addition of passive reflecting elements without necessarily increasing the number of RF chains at the transmitter \cite{Jamali.2021}, which has been exploited to design more spatially flexible single-RF chain transmitters \cite{Bereyhi.2020,Rosabal.2024}. 

\subsection{Motivation and Contributions}\label{subsec:motivationsAndContributions}
Only a few studies, \textit{e.g.}, \cite{Rosabal.2024}, have explored the potential of RIS-equipped transmitters in WET-enabled settings, which motivates our work. In fact, the focus of \cite{Samaiyar.2023,Jamali.2021,Bereyhi.2020} is on wireless information transmission and therefore the obtained insights are not extensible to WET. Unlike \cite{Rosabal.2024}, we adopt an intelligent transmitting (refracting) surface (ITS) to avoid the blockage problems caused by the feeder structure, albeit at the expense of a slight increase in insertion losses at the surface. Notably, the procedures herein presented are extensible to the architecture in which the surface operates in the reflecting mode and only the performance results may differ.

In this work, we aim at minimizing the power consumption of an ITS-equipped PB, with multiple RF chains, charging a network of single-antenna IoT devices. To this end, we formulate a jointly optimization problem over the ITS's phase shifter configuration and the feeders' digital precoder, under the practical assumption that each RF chain is terminated with a Doherty high-power amplifier (HPA). Therefore, our problem accounts for the intrinsic nonlinearities of the HPAs, the constant modulus constraints on the passive phase shifters, and the highly nonlinear minimum received power at the devices. Our contributions are as follows: 
\begin{itemize}
    \item We characterize the power consumption of the ITS-equipped PB using multiple RF chains. To this end, we account for the nonlinearities of the HPA driving the feeder antennas, the control power required to operate the ITS, and the inherent losses in the air interface between the feeder and the ITS.
    \item We formulate a nonconvex power minimization problem to satisfy a target received power at the IoT devices. To tackle the problem's  complexity, we transform the objective and constraints to ensure efficient application of the successive convex approximation (SCA) framework. This allows us to re-cast the original problem into a series of convex subproblems jointly optimizing the feeders' digital precoder and the ITS phase shifter configuration.
    \item We propose an initialization algorithm to address the sensitivity of the SCA method to the initial guess. Such procedure divides the PB into single-RF chain virtual transmitters, each illuminating a distinct, nonoverlapping subset of the ITS. It then explores different virtual transmitter-to-device associations and selects the one that minimizes the power consumption of the HPAs. Results demonstrate the effectiveness of the proposed algorithm in balancing convergence speed and solution quality. 
    \item We illustrate improvements in complexity and power consumption of the ITS-equipped PB compared with digital and hybrid analog-digital beamforming implementations. Results confirm that the ITS-equipped PB architecture achieves superior performance across different hardware configurations, highlighting its effectiveness as a practical solution for implementing energy-efficient beamforming. 
    \item We show the impact of the geometry of the PB on the channel propagation conditions. Specifically, we illustrate that nonuniform power distribution across the ITS can cause a device to operate in the near- or far-field region, even when the ITS has the same physical aperture.
\end{itemize}
\subsection{Outline}
Next, Section~\ref{sec:systemModel} introduces the system model and presents the problem formulation. Section~\ref{sec:powerConsumptionModel} follows with a discussion of the power consumption model and the main sources of losses. Then, Section~\ref{sec:proposedSolution} outlines the proposed optimization framework, while Section~\ref{sec:benchmarkSolutions} elaborates on benchmark beamforming technologies. Subsequently, Section~\ref{sec:numericalAnalysis} analyzes the performance of the proposed PB architecture against the benchmark technologies. Finally, Section~\ref{sec:conclusions} summarizes the findings and concludes this work.


\subsection{Notation}
\begin{table}[t]
    \centering
    \caption{List of abbreviations and symbols}
    \label{tab:sym}
    \begin{tabular}{p{1.2cm} p{7cm}}
        \thickhline
            \textbf{Parameter} & \textbf{Definition} \\
        \hline
            FD   & fully-digital beamforming \\
            HBFC & hybrid analog-digital beamforming (fully-connected) \\
            HBPC & hybrid analog-digital beamforming (partially-connected) \\
            HPA  & high-power amplifier \\
            IoT  & Internet of Things \\
            ITS  & intelligent transmitting surface \\
            PB   & power beacon \\
            RF   & radio-frequency \\
            SCA  & successive convex approximation \\
            SDP  & semidefinite program \\
            SOCP & second-order cone program \\
            WET  & wireless energy transfer \\
            \hline
            $\mathcal{N}$ & set of $N$ antennas \\
            $\mathcal{M}$ & set of $M$ ITS elements \\
            $\mathcal{K}$ & set of $K$ IoT devices \\
            $Q$ & number of energy-carrying signals \\
            $\bm{\phi}$ & ITS configuration vector \\
            $\mathbf{s}$ & baseband signal vector \\
            $\mathbf{b}_q$ & digital precoder for symbol $s_q$ \\
            $g$ & HPA's power gain \\
            $\ell$ & number of HPA in the Doherty architecture \\
            $\eta_\mathrm{max}$ & Doherty HPA's maximum efficiency \\
            $P_\mathrm{max}$ & Doherty HPA's maximum output power \\
            $P^{i}_T$ & PB's power consumption under architecture $i \in \{\mathrm{its}, \mathrm{fd}, \mathrm{hb}\}$ \\
            $P_n, P_{\mathrm{hpa},n}$ & $n^\mathrm{th}$ HPA's output and consumed power, respectively \\
            $P_\mathrm{ctrl}$ & ITS's control power \\
            $P_\mathrm{cell}$ & ITS's passive element power consumption \\
            $\rho_\mathrm{its}$ & ITS's power efficiency factor \\
            $\mathbf{h}_k$ & complex channel vector ($k^\mathrm{th}$ device to ITS) \\
            $\mathbf{A}$ & complex channel matrix (feeder-to-ITS) \\
            $P_k$ & received power at the $k^\mathrm{th}$ device \\
            $F(\beta,\xi)$ & antenna's radiation profile at boresight angle $\beta$ with gain $\xi$ \\
            $\theta_{m,n}$ & ITS angle from boresight (relative to the feeder) \\
            $\vartheta_{n,m}$ & feeder angle from boresight (relative to the ITS) \\
            $\zeta_{k,m}$ & device angle from boresight (relative to the ITS) \\
            $\kappa, \mu$ & boresight gains of ITS's elements and antennas, respectively \\
            $\lambda$ & wavelenght of the transmitted signal \\
            $\mathbf{v}_n$ & Cartesian coordinates of the $n^\mathrm{th}$ feeder's antenna \\
            $\mathbf{u}_k$ & Cartesian coordinates of the $k^\mathrm{th}$ device \\
            $\mathbf{r}_m$ & Cartesian coordinates of the $m^\mathrm{th}$ ITS element \\
            $\delta_a$ & distance ITS to the feeder \\
            $r_a$ & radius of the feeder \\
        \thickhline
    \end{tabular}
\end{table} 
We use boldface lowercase letters to denote column vectors and boldface uppercase letters to denote matrices. $\lVert \mathbf{x} \rVert_2$ denotes the $\ell_2$-norm of $\mathbf{x}$, while $(\cdot)^T$, $(\cdot)^H$, $|\cdot|$, $\lfloor \cdot \rfloor$, and $\mathrm{mod}(\cdot)$ denote the transpose, Hermitian transpose, absolute value, floor, and modulus operations, respectively. Also, $\mathbb{C}$ is the set of complex numbers with imaginary unit $j = \sqrt{-1}$, and $\mathbb{E}_\mathbf{s}[\cdot]$ is the expectation operator concerning the random vector $\mathbf{s}$. Moreover, $\mathbf{I}_N$, $\mathrm{tr}(\cdot)$, $\setminus$, and $\angle$ denote an $N-$dimensional identity matrix, the trace operator, the set difference operator, and the phase of a complex number, respectively. We utilize $x^*$ to denote the final solution of a problem with optimization variable $x$ and $\Tilde{x}$ to denote an intermediate feasible guess of the problem solution when applying SCA. Also, $\mathcal{O}(\cdot)$ is the big-O notation, which specifies worst-case complexity. Finally, the $\mathrm{blkdiag}(\mathbf{c}_1, \ \mathbf{c}_2, \ \ldots, \ \mathbf{c}_n, \ \ldots, \ \mathbf{c}_N)$ denotes the block-diagonal concatenation of the column vectors $\{\mathbf{c}_n\}$. 

\section{System Model}\label{sec:systemModel}
\begin{figure}[t!]
    \centering
    \includegraphics[width=\linewidth]{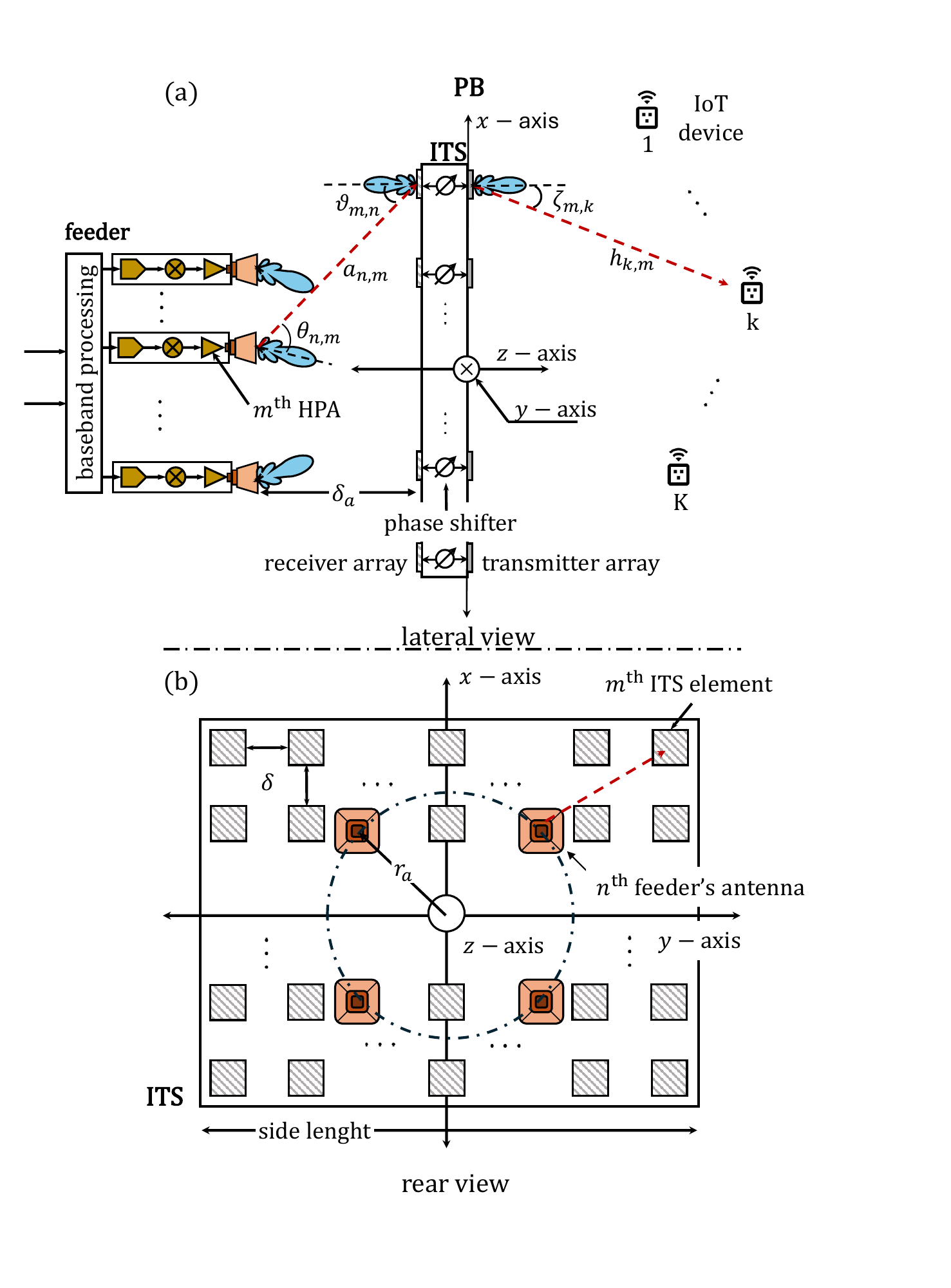}
    \vspace{-2em}
    \caption{ITS-assisted PB charges $K$ single-antenna IoT devices. A collocated digital beamforming transmitter illuminates the receiving side of the ITS which then retransmits a phase-shifted version of the received signals. The receiving and transmitting sides of the ITS are connected via a set of passive phase-shifting elements. We consider that the centroid of the feeder's array is aligned with the center of the ITS.}
    \label{fig:systemModel}
\end{figure}
 Consider the scenario illustrated in Fig.~\ref{fig:systemModel}, where an ITS-equipped PB charges a set $\mathcal{K}$ of $K$ single-antenna IoT devices. The PB architecture comprises a passive ITS with a set $\mathcal{M}$ of $M$ elements and a physically close digital beamforming-based feeder with a set $\mathcal{N}$ of $N \geq K$ antennas connected to dedicated HPAs. The ITS comprises two planar antenna arrays connected by phase-shifting elements, with one side acting as a receiver and the other as a transmitter \cite{Abdelrahman.2017}. Moreover, $\mathbf{u}_k \in \mathbb{R}^{3\times 1}$, $\mathbf{r}_m \in \mathbb{R}^{3\times 1}$, and $\mathbf{v}_n \in \mathbb{R}^{3\times 1}$ denote the Cartesian coordinates of the $k^\mathrm{th}$ device, the $m^\mathrm{th}$ ITS passive element, and $n^\mathrm{th}$ feeder's antenna, respectively. $\mathbf{A} \in \mathbb{C}^{M \times N}$ denotes the matrix of channel coefficients between the feeder antennas and the ITS elements. For instance, the entry $a_{n,m}$ of $\mathbf{A}$ is the channel coefficient between the $n^\mathrm{th}$ feeder's antenna and the $m^\mathrm{th}$ ITS element. Similarly, $\mathbf{h}_k \in \mathbb{C}^{M \times 1}$, with entries $\{h_{k,m}\}$, is the channel vector between the ITS and the $k^\mathrm{th}$ device. 

 \subsection{Signal model}\label{subsec:signalModel}
 The PB transmits the baseband signal vector $\mathbf{s} \in \mathbb{C}^{Q \times 1}$, with $Q \leq N$ independent symbols $\{s_q\}$, \textit{i.e.}, $\mathbb{E}[\mathbf{s}\mathbf{s}^H] = \mathbf{I}_Q$, to charge the IoT devices. The corresponding set of $Q$ digital precoders is $\{\mathbf{b}_q\}$ with $\mathbf{b}_q \in \mathbb{C}^{N \times 1}$. Observe that the condition $Q \leq N$ accounts for scenarios where favorable channel conditions enable the devices to be charged with fewer energy-carrying signals than the total number of physical RF chains available on the PB. We neglect the impact of the feeder direct transmission on the signal received at the IoT devices, as the latter is (intended to be) dominated by the ITS transmission. In fact, its impact, if any, would be for IoT devices at specific locations (\textit{i.e.}, close and in the direction of the feeder's boresight direction). 
 
 The received signal at the $k^\mathrm{th}$ device ignoring the noise power is 
{\setlength{\abovedisplayskip}{4pt}
 \setlength{\belowdisplayskip}{4pt}
\begin{equation}
    y_k = \mathbf{h}^H_k\mathrm{diag}(\bm{\phi})\mathbf{A} \sum_{q = 1}^{Q} \sqrt{g\rho_\mathrm{its}}\mathbf{b}_q s_q,
\end{equation}
}
where $\bm{\phi} \in \mathbb{C}^{M\times1}$ is the ITS phase shift configuration vector, whose $m^\mathrm{th}$ entry is $\phi_m = e^{j\beta_m}$ with $\beta_m \in [0, 2\pi]$ and $g$ is the power gain of the HPA. Moreover, $\rho_\mathrm{its}$ denotes the ITS power efficiency factor, which is explained in detail in Section~\ref{subsec:Losses}. Therefore, the received power at the $k^\mathrm{th}$ device is given by
{\setlength{\abovedisplayskip}{4pt}
 \setlength{\belowdisplayskip}{8pt}
\begin{align}
    P_k &= \mathbb{E}_\mathbf{s}\left[|y_k|^2\right] = \mathbb{E}_\mathbf{s}\left[\left| \mathbf{h}^H_k\mathrm{diag}(\bm{\phi})\mathbf{A} \sum_{q = 1}^{Q} \sqrt{g\rho_\mathrm{its}}\mathbf{b}_q s_q\right|^2\right] \nonumber \\
    &= g\rho_\mathrm{irs} \sum_{q=1}^{Q}|\mathbf{h}^H_k \mathrm{diag}(\bm{\phi}) \mathbf{A} \mathbf{b}_q|^2.
\end{align}
}
\subsection{Channel model}\label{sub:channelModel}
We assume that both the feeder and the devices operate in the near-field region of the ITS; hence, the entries of $\mathbf{A}$ and $\mathbf{h}_{k}$ are computed as \cite{Zhang.2022}
\begin{align}
    a_{n,m} &= \sqrt{F(\theta_{m,n},\mu)F(\vartheta_{n,m},\kappa)}\frac{\lambda e^{-2\pi j \lVert \mathbf{v}_n - \mathbf{r}_m \rVert_2/\lambda}}{4\pi\lVert \mathbf{v}_n - \mathbf{r}_m \rVert_2}, \\
    h_{k,m} &= \sqrt{F(\zeta_{k,m},\kappa)}\frac{\lambda e^{-2\pi j \lVert \mathbf{u}_k - \mathbf{r}_m \rVert_2/\lambda}}{4\pi\lVert \mathbf{u}_k - \mathbf{r}_m \lVert_2}, \label{eq:channelITSToDevice}
\end{align}
where, $\theta_{m,n}$ is the angle from boresight of the $m^\mathrm{th}$ passive element measured with respect to the $n^\mathrm{th}$ feeder's antenna. Similarly, $\vartheta_{n,m}$ and $\zeta_{k,m}$ are the angles from boresight of the $n^\mathrm{th}$ feeder's antenna and the $k^\mathrm{th}$ devices, respectively, measured from the $m^\mathrm{th}$ ITS passive element. Moreover, $\kappa$ and $\mu$ (both $\geq 2$) are the boresight gains of the passive elements and the feeder's antennas, respectively, and
\begin{equation}
    F(\beta,\xi) = \begin{cases}
        2(\xi+1)\cos^\xi(\beta), \ &\beta \in \left[0,\frac{\pi}{2}\right] \\
        0, \ &\text{otherwise}
        \end{cases}
\end{equation}
is the radiation profile of the antenna elements. It is worth mentioning that the channel model presented here neglects any coupling between the feeder and the ITS arising from their close physical proximity. Accounting for these effects would necessitate an alternative modeling approach. Consequently, we select the simulation parameters in Section~\ref{sec:numericalAnalysis} to reduce the impact of mutual coupling on the results.

Finally, we consider a full illumination strategy in which the radiation patterns of the feeder's antennas are directed to the center of the ITS, as shown in Fig.~\ref{fig:systemModel}.a. The feeder's antennas are placed in the vertices of a regular polygon whose circumscribed circumference has radius $r_a$ (see Fig.~\ref{fig:systemModel}.b) and its center is aligned to that of the ITS at a distance $\delta_a$ (see Fig.~\ref{fig:systemModel}.a). Adjacent elements, as shown in Fig.~\ref{fig:systemModel}b in the ITS are spaced $\delta$ units apart along both the $x-$ and $y-$axes. A more detailed discussion of alternative illumination strategies can be found in \cite{Jamali.2021}.

\subsection{Aperture losses}\label{subsec:Losses}
Since the ITS is finite, some of the power radiated by the feeder will not be captured by the passive antennas, resulting in spillover losses. These losses are implicitly captured by the channel matrix $\mathbf{A}$ along with the received power variations in the ITS surface. The amount of power loss depends on the illumination strategy, the distance between the feeder and the ITS, the radiation pattern of the antenna elements, and the area of the ITS. Moreover, in practice, the actual power radiated into the wireless channel is smaller than the one captured by the surface due to inefficiencies in the antennas, unwanted reflections from the receiving side of the ITS, and power absorption in the ITS \cite{Jamali.2021}, which are captured by the power efficiency factor $\rho_\mathrm{its}$. Finally, each phase shifter introduces a certain loss when the signal travels from a given antenna element on the receiving side to the corresponding one on the transmitting side. Although one can attribute this loss to the specific value of phase shift, in most modern implementations it can be assumed independent of the phase shifter configuration \cite{Wang.2024}.

\section{Power consumption model}\label{sec:powerConsumptionModel}
We model the total power consumption of the PB as
\begin{equation}\label{eq:itsPowConsumptionModel}
    P^\mathrm{its}_T = P_\mathrm{bb} + P_\mathrm{tc} + P_\mathrm{hpa} + P_\mathrm{its},
\end{equation}
where $P_\mathrm{bb}$, $P_\mathrm{tc}$, $P_\mathrm{hpa}$, and $P_\mathrm{its}$ denote the power consumption of the baseband circuitry, the transceiver chains, the HPAs, and the passive ITS, respectively.

\subsection{Baseband circuitry power consumption} 
The baseband circuit power consumption comprises the power consumed for digital baseband processing $P_\mathrm{bb}$ and the transceiver chains $P_\mathrm{tc}$. The operations carried out by the digital processing block include channel estimation, precoder computation, and digital predistortion techniques to compensate for the nonlinearities of the PB's circuits. The main factors impacting baseband digital processing include signal bandwidth, the number of transmit antennas, and the efficiency of the applied signal processing algorithms (see for instance \cite{Bjornson.2015} and references therein for more details). Additionally, $P_\mathrm{tc}$ is the power required to run the circuit components including converters, mixers, and filters that process the feeding signals to the antennas in the analog domain. Moreover, this term also accounts for the power consumption of the local oscillator. Although these two components of the PB's power consumption may vary with the operation conditions, we assume herein that they remain constant \cite{Bjornson.2015,Jamali.2021}.

\subsection{HPA power consumption}\label{subsec:HPAPowerConsumption}
\begin{figure}[t]
    \centering
    \includegraphics[width=\linewidth]{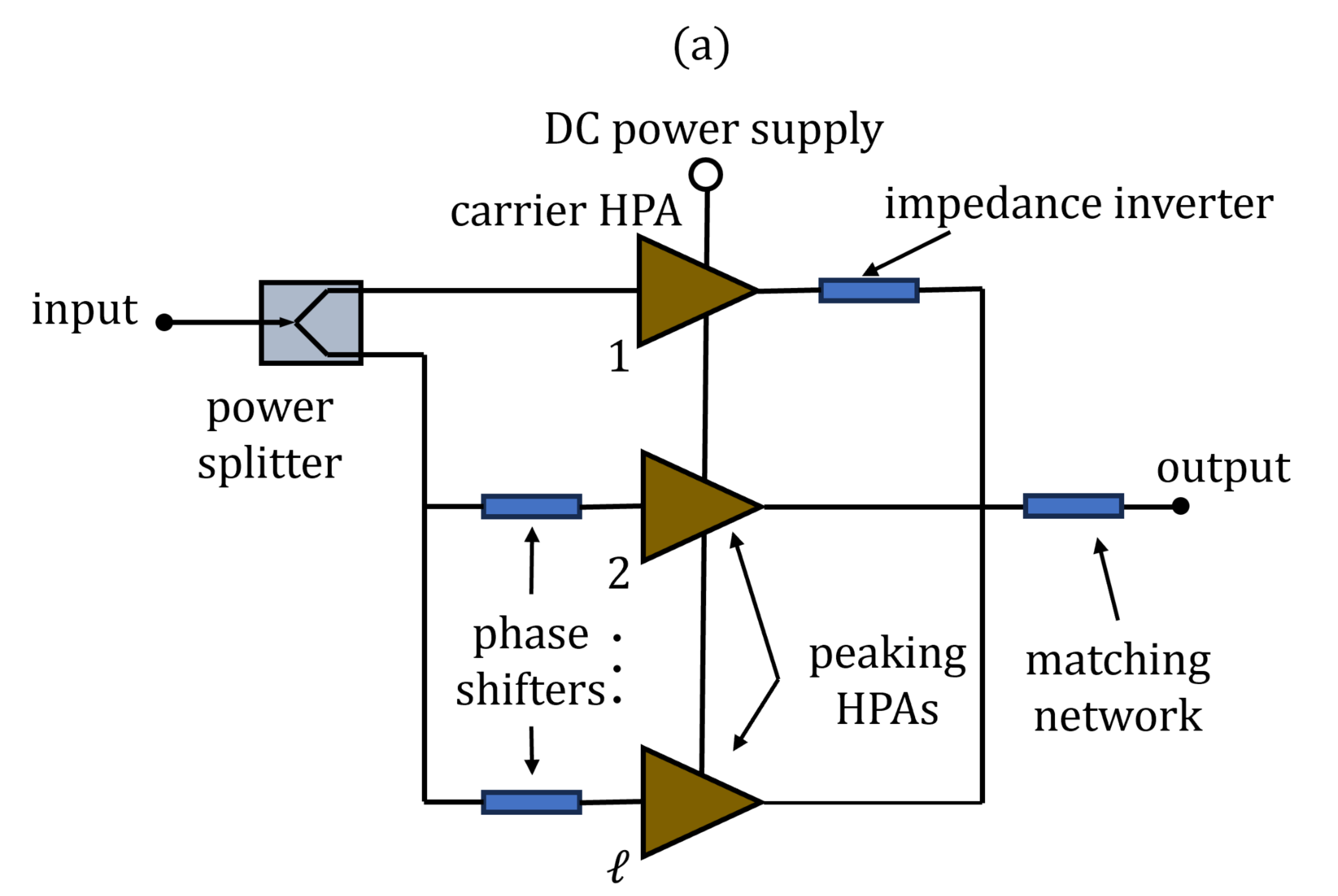}
    \\
    \vspace{1em}
    \includegraphics[width=\linewidth]{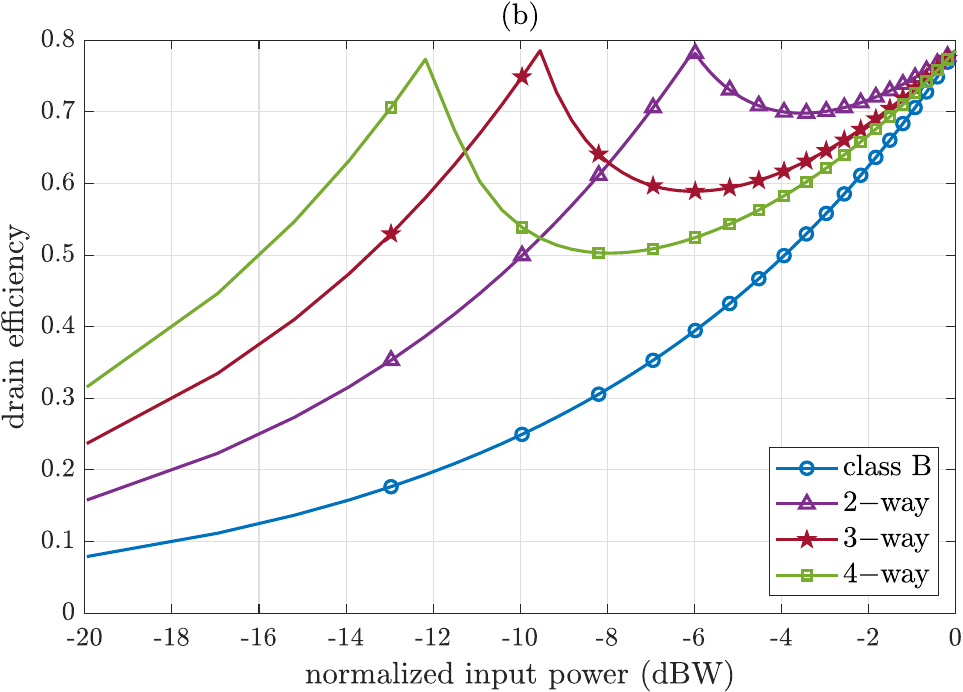}
    \vspace{-1.5em}
    \caption{$\ell-$way Doherty HPA: (a) symmetric architecture and (b) drain efficiency versus the normalized input power. In (a), the blue rectangles indicate blocks implemented using transmission lines whereas the dark orange triangles represent HPAs. The power splitter feeds $\ell - 1$ peaking HPAs once the input signal exceeds a fraction $\frac{1}{\ell}$ of the total signal voltage.}
    \label{fig:DohertyAmplifier}
\end{figure}

Each RF chain employs an $\ell-$way Doherty HPA comprising one carrier and $\ell - 1$ peaking HPAs, as depicted in Fig.~\ref{fig:DohertyAmplifier}. Doherty HPAs achieve high power efficiency across a wide input power range using a relatively cost-effective architecture \cite{Nikandish.2020}. Under normal operation, only the carrier HPA is active; however, if the input signal voltage surpasses $\frac{1}{\ell}$ of its peak value, a power-dividing network activates the peaking HPAs to provide additional amplification. The amplified signals are then coherently combined at the amplifier's output. Such behavior is particularly beneficial in WET systems, where signals with high peak-to-average power ratios enhance the RF-to-DC conversion efficiency of RF-EH circuits \cite{Clerckx.2021,Collado.2014}.

Back to our scenario, the output signal at the $n^\mathrm{th}$ HPA is $\sum_{q=1}^{Q} \sqrt{g} b_{q,n}  s_q$, and hence its output power is given by
\begin{equation}\label{eq:inputPowDoherty}
    P_n = \mathbb{E}_s\left[ \left| \sqrt{g}\sum_{q = 1}^{Q} b_{q,n}  s_q \right|^2 \right] =  g\sum_{q=1}^{Q} |b_{q,n}|^2,
\end{equation}
where the final step follows from the fact that the expectations of the cross terms in the expansion of the square absolute value operation are zero, \textit{i.e.}, $\mathbb{E}[\mathbf{s}\mathbf{s}^H] = \mathbf{I}_Q$. Considering the working principle previously described, herein, we adopt a piecewise function to model the power consumption of the $n^\mathrm{th}$ HPA as \cite{Joung.2014,Lin.2016}
\begin{equation}\label{eq:dohertyPowConsumpt}
    P_{\mathrm{hpa},n} = \begin{dcases}
        \frac{\sqrt{P_n P_\mathrm{max}}}{\ell \eta_\mathrm{max}}, &0 < P_n \leq \frac{P_\mathrm{max}}{\ell^2}, \\
        \frac{(\ell + 1)\sqrt{P_n P_\mathrm{max}}-P_\mathrm{max}}{\ell \eta_\mathrm{max}}, &\frac{P_\mathrm{max}}{\ell^2} < P_n \leq P_\mathrm{max},
        \end{dcases}
\end{equation}
where $P_\mathrm{max}$ is the maximum output power and $\eta_\mathrm{max}$ is the maximum efficiency achievable by each Doherty HPA. The threshold at $\frac{P_\mathrm{max}}{\ell^2}$ marks the transition between two regions: the upper region, where only the carrier HPA operates, and the lower region, where the peaking HPAs are activated. Finally, the total HPA power consumption is given by $P_\mathrm{hpa} = \sum_{n=1}^{N} P_{\mathrm{hpa},n}$.

Fig.~\ref{fig:DohertyAmplifier} illustrates the drain efficiency---defined as the ratio between the HPA's output power and its corresponding power consumption---of an $\ell-$way symmetric Doherty amplifier for different values of the (normalized) input signal and different values of the parameter $\ell$ considering the power consumption model in \eqref{eq:dohertyPowConsumpt}. The fficiency improves as the number of peaking HPAs increases because each amplifier handles a smaller fraction of the input voltage. Since all peaking HPAs---implemented here as class B amplifiers---activate simultaneously, only a single back-off point is observed. Despite these benefits, the process of designing and tuning the parameters of Doherty HPAs is more involved than that associated to its constituent HPAs blocks when operated individually \cite{Choi.2022,Chen.2023}. We refer the reader to \cite{Piacibello.2023,Li.2020} for more detailed information about other Doherty HPA implementations.

\subsection{ITS power consumption}\label{subsec:ITSPowerConsumption}
The power consumption of the passive ITS can be attributed to the control board, the drive circuit, and the phase-shifting elements \cite{Wang.2024,Wang2.2023}. The power consumption of the control board is typically assumed to be constant, whereas that of the driver circuits varies depending on the specific circuit requirements for each passive element technology, the number of control signals needed, the number of passive elements sharing a common control signal, and the capacity of each driver to generate control signals \cite{Wang.2023}. Meanwhile, the power consumption of the passive reflective elements depends on the particular technology used to implement them which can be positive-intrinsic-negative (PIN) diodes, varactor diodes, and RF switches. For instance, the power consumption when using PIN diodes is affected by the number of required diodes to achieve different polarization modes, the resolution of each passive element, and the number of on-state bits in the coding state of the elements \cite{Wang.2024}. Implementations based on RF switches require the consideration of the polarization mode, number of cells, and inherent power consumption of individual switches \cite{Wang.2024}. In contrast, when using varactor diodes, the power consumption is negligible even for many elements \cite{Pei.2021}.

Herein, we model the power consumption of the ITS as 
\begin{equation}
    P_\mathrm{its} = P_\mathrm{ctrl} + MP_\mathrm{cell},
\end{equation}
where $P_\mathrm{ctrl}$ captures the power consumption of the control board and that of the drive circuits, whereas $P_\mathrm{cell}$ denotes the power consumption of the passive reflective element.

\section{Problem formulation and proposed optimization framework}\label{sec:proposedSolution}
In this section, we leverage the power consumption model discussed earlier to formulate our main optimization problem and introduce a sequence of transformations that enable its efficient solution via SCA.

\subsection{Problem formulation}\label{subsec:problemFormulation}
We aim to minimize the PB's total power consumption $P^\mathrm{its}_\mathrm{T}$ while delivering a target RF power to the IoT devices, denoted as $P^\mathrm{th}_k$ for device $k$. To this end, we formulate an optimization problem as follows
\begin{subequations}\label{P1}
    \begin{alignat}{2}
    \mathbf{P1:} \ &\underset{\{\mathbf{b}_q\},\bm{\phi}}{\mathrm{minimize}} \ && P^\mathrm{its}_\mathrm{T} \label{P1a}\\
    &\text{s.t.} \ && P_k \geq P_k^\mathrm{th} , \ \forall k \in \cal K, \label{P1b}\\
    & \ && |\phi_m| = 1, \ m = 1, \ldots, M,\label{P1c}
    \end{alignat}
\end{subequations}
where \eqref{P1b} ensures a minimum receive power at all devices and \eqref{P1c} enforces unit modulus of the phase shift amplitudes, reflecting the passive nature of the ITS elements. Observe that the PB's maximum transmit power is indirectly controlled by the parameter $P_\mathrm{max}$ in \eqref{eq:dohertyPowConsumpt}, which constrains the total transmit power through the weighted sum of the beamformed signals at the input of each HPA. Notably, $\mathbf{P1}$ is nonconvex since the unit-modulus constraints \eqref{P1c} define a discrete set, while neither the received power constraints \eqref{P1b} nor the objective function are convex.

\subsection{Problem reformulation}\label{subsec:problemReformulation}
Let us rewrite the objective function $\mathbf{P1}$ in epigraph form \cite{Boyd.2004}
\begin{subequations}\label{eq:epigraph}
    \begin{alignat}{2}
    & \underset{\{\mathbf{b}_q\},\bm{\phi},\{t_n\}}{\mathrm{minimize}} \ && \sum_{n=1}^{N} t_n\label{eq:epigrapha} \\
    & \ \text{s.t.} \ && P_{\mathrm{hpa},n} \leq t_n, \ \forall n \in \mathcal{N}, \label{eq:epigraphb} \\
    & \ && \ t_n \geq 0, \ \forall n \in \mathcal{N},  \label{eq:epigraphc} 
    \end{alignat}
\end{subequations}
where $\{t_n\}$ are the auxiliary variables. Herein, we have ignored the terms in \eqref{eq:itsPowConsumptionModel} that are independent of the optimization variables since they have no impact on how to reach the optimal point. However, these are latter considered in Section~\ref{sec:numericalAnalysis} as they are essential to assess the PB total power consumption. Notice that $P_{\mathrm{hpa},n}$ is defined as a piecewise function which stems from the fact that we are considering a multistage HPA. To overcome this problem, let $\{\Tilde{\mathbf{b}}_q\}$ denote feasible precoding vectors satisfying the constraints \eqref{P1b} and the maximum operating power for the HPAs. An appropriate subfunction of \eqref{eq:dohertyPowConsumpt} can then be selected based on the output power corresponding to these initial precoding vectors. This procedure is not required for a single-stage HPA resulting in a simpler reformulation of the objective. However, as shown in Fig.~\ref{fig:DohertyAmplifier}b, the corresponding drain efficiency is lower thus resulting in a higher consumed power. Notably, the proposed procedure narrows the search space because only a subset of the constraints become active, hence enabling a simpler and more tractable solution process of $\mathbf{P1}$, while still respecting the key power constraints. Moreover, since the subfunctions are convex and nondecreasing, optimality is not compromised when applying this procedure \cite{Duchi.2018}. For instance, when
\begin{equation}\label{eq:cond1}
    g\sum_{q=1}^Q|\Tilde{b}_{q,n}|^2 \leq \frac{P_\mathrm{max}}{\ell^2},
\end{equation}
meaning that only the carrier amplifier is active according to \eqref{eq:dohertyPowConsumpt}, constraints \eqref{eq:epigraphb} transform into
\begin{subequations}
    \begin{align}
        \frac{\sqrt{g\sum_{q=1}^Q|b_{q,n}|^2 P_\mathrm{max}}}{\ell \eta_\mathrm{max}}&\leq t_n, \label{eq:branch1.1} \\
        g\sum_{q=1}^Q|b_{q,n}|^2 &\leq \frac{P_\mathrm{max}}{\ell^2} \label{eq:branch1.2}.
    \end{align}
\end{subequations}
which are convex. Similarly, if
\begin{equation}\label{eq:cond2}
    \frac{P_\mathrm{max}}{\ell^2} < g\sum_{q=1}^Q|\Tilde{b}_{q,n}|^2 \leq P_\mathrm{max},
\end{equation}
meaning that the peaking HPAs are also active, constraints \eqref{eq:epigraphb} transform into
\begin{subequations}
    \begin{align}
        \frac{(\ell + 1)\sqrt{g\sum_{q=1}^Q|b_{q,n}|^2 P_\mathrm{max}}-P_\mathrm{max}}{\ell \eta_\mathrm{max}} &\leq t_n, \label{eq:branch2.1} \\
        g\sum_{q=1}^Q|b_{q,n}|^2 &\leq P_\mathrm{max}, \label{eq:branch2.2} \\
        g\sum_{q=1}^Q|b_{q,n}|^2 &> \frac{P_\mathrm{max}}{\ell^2}. \label{eq:branch2.3}
    \end{align}
\end{subequations}
Since \eqref{eq:branch2.3} is not convex, we construct a convex surrogate function using its first-order Taylor approximation as
\begin{equation}\label{eq:branch2.3approx}
    g\sum_{q=1}^{Q} 2\Re\left\{\Tilde{b}_{q,n}^Hb_{q,n}\right\} + |\Tilde{b}_{q,n}|^2 \ge \frac{P_\mathrm{max}}{\ell^2}.
\end{equation}

We now deal with constraints \eqref{P1b}. Let us define $\hat{\mathbf{h}}^H_k = \mathbf{h}^H_k\mathrm{diag}(\bm{\phi})\mathbf{A}$ and find a concave lower bound for the right-hand side of \eqref{P1b} in the form
\begin{equation}    g\rho_\mathrm{its}\sum_{q=1}^{Q}|\hat{\mathbf{h}}^H_k \mathbf{b}_q|^2 \ge g\rho_\mathrm{its} \sum_{q=1}^{Q}\Tilde{P}_{k,q}. 
\end{equation}
Inspired by \cite{Kumar.2023}, we lower bound each addend $|\Tilde{\mathbf{h}}^H_k \mathbf{b}_q|^2$ as
\begin{align} \label{eq:rPowApprox}
    |\Tilde{\mathbf{h}}^H_k \mathbf{b}_q|^2 &\stackrel{(a)}{\geq} 2\Re\left\{z^H_{k,q}\hat{\mathbf{h}}^H_k \mathbf{b}_q\right\} - |z_{k,q}|^2 \nonumber \\
    &\stackrel{(b)}{=} \frac{1}{2} \left(\lVert z_{k,q}\hat{\mathbf{h}}_k + \mathbf{b}_q \rVert^2 - \lVert z_{k,q}\hat{\mathbf{h}}_k - \mathbf{b}_q \rVert^2\right) - |z_{k,q}|^2 \nonumber \\
    &\stackrel{(c)}{\geq} \Re\left\{\bm{\nu}^H_{k,q}\left(z_{k,q} \hat{\mathbf{h}}_k + \mathbf{b}_q\right)\right\} - \frac{1}{2}\lVert \bm{\nu}_{k,q} \rVert^2 - \nonumber \\
    & \qquad \frac{1}{2}\lVert z_{k,q}\hat{\mathbf{h}}_k - \mathbf{b}_q \rVert^2 - |z_{k,q}|^2 \nonumber \\ 
    &\triangleq \Tilde{P}_{k,q},
\end{align}
where (a) is the first-order approximation of $|\hat{\mathbf{h}}^H_k \mathbf{b}_q|^2$ around the point $z_{k,q} = \Tilde{\mathbf{h}}^H_k \Tilde{\mathbf{b}}_q$, (b) uses the identity $\Re\{\mathbf{x}^H\mathbf{y}\} = \frac{1}{4}\left[\lVert \mathbf{x} + \mathbf{y} \rVert^2 - \lVert \mathbf{x} - \mathbf{y} \rVert^2 \right]$ and (c) contains a first-order approximation of the term $\lVert z_{k,q}\Tilde{\mathbf{h}}_k + \mathbf{b}_q \rVert^2$ around the point $\bm{\nu}_{k,q} = z_{k,q}\Tilde{\mathbf{h}}_k + \Tilde{\mathbf{b}}_q$. Herein, $\Tilde{\mathbf{h}}^H_k = \mathbf{h}^H_k\mathrm{diag}(\Tilde{\bm{\phi}})\mathbf{A}$ is a function of a feasible phase shift configuration. We note that when utilizing the SCA framework one can restrict the solution space of the optimization variables to ensure high accuracy of the approximation around the current iterate for which the first-order approximation holds tight \cite{Duchi.2018}. Finally, we relax the constant modulus constraints \eqref{P1c} as $|\phi_m| \leq 1$, which will hold with equality at the optimal point. This is because, different from communication setups in which interference management is required, in WET-enabled networks, a solution with $|\phi_m| < 1$ will attenuate the received signal and therefore reduce the devices' harvested energy. We can now reformulate $\mathbf{P1}$ as the following convex problem
\begin{subequations}\label{P2}
    \begin{alignat}{2}
    \mathbf{P2:} \ &\underset{\{\mathbf{b}_q\},\bm{\phi}, \{t_n\}}{\mathrm{minimize}} \ && \sum_{n=1}^{N} t_n \label{P2a}\\
    &\text{s.t.} && \eqref{eq:branch1.1},\eqref{eq:branch1.2}, \ \{\forall n \in \mathcal{N} \ | \ \eqref{eq:cond1} \}, \label{P2b} \\ 
    & \ && \eqref{eq:branch2.1},\eqref{eq:branch2.2},\eqref{eq:branch2.3approx}, \ \{\forall n \in \mathcal{N} \ | \ \eqref{eq:cond2} \}, \label{P2c} \\
    & \ && g\rho_\mathrm{its} \sum_{q=1}^{Q}\Tilde{P}_k \geq P_k^\mathrm{th} , \ \forall k \in \cal K, \label{P2d}\\
    & \ && |\phi_m| \leq 1, \ \forall m \in \mathcal{M}, \label{P2e} \\
    & \ && t_n \geq 0, \ \forall n \in \mathcal{N}. \label{P2f} 
    \end{alignat}
\end{subequations}
Depending on the values $\{\Tilde{b}_{q}\}$, either constraint \eqref{P2b} or \eqref{P2c} is activated, but they are never enforced simultaneously for a single RF chain.

To obtain a solution of $\mathbf{P1}$, we solve a sequence of subproblems in the form of $\mathbf{P2}$, using given values of $\{\Tilde{\mathbf{b}}_q\}$ and $\bm{\Tilde{\phi}}$, as outlined in Algorithm~\ref{alg:SCA}. After each iteration, we update our estimates for the optimal precoders and phase shifter configurations using the solution of $\mathbf{P2}$. The algorithm is considered to have converged when the changes in the objective function \eqref{P2a} become negligible. The most critical step in the proposed solution requires solving a second-order cone program (SOCP) reformulation of $\mathbf{P2}$ with $4N + M + K$ constraints and $2N+M$ variables, which for $M \gg N$ necessitates $\mathcal{O}(\sqrt{M})$ iterations, each costing $\mathcal{O}(M^{3.5})$ arithmetic operations when using the interior-point method (IPM) \cite[Section 6.6.2]{Ben.2001}. Note that Algorithm~\ref{alg:SCA} relies on the knowledge of an initial phase shift configuration and digital precoders, which is discussed next. 
\begin{algorithm}[t!]
\caption{SCA-based solution}
\begin{algorithmic}[1] \label{alg:SCA}
 \STATE \textbf{Input:} $\{P^\mathrm{th}_k, \mathbf{h}_k\}_{\forall k\in\mathcal{K}}, \mathbf{A}, N, M, \Tilde{\bm{\phi}}, \{\Tilde{{\mathbf{b}}}_q\}$ \label{alg:SCA1}
\REPEAT
\STATE Update $z_{k,q} = \Tilde{\mathbf{h}}^H_k \Tilde{\mathbf{b}}_q$ and $\bm{\nu}_{k,q} = z_{k,q} \Tilde{\mathbf{h}}_k + \Tilde{\mathbf{b}}_q$\label{alg:SCA3}
\STATE Solve $\mathbf{P2}$ \label{alg:SCA4}
\STATE Update: $\Tilde{\mathbf{b}}_q \gets \mathbf{b}^*_q$ and $\Tilde{\bm{\phi}} \gets \bm{\phi}^*$
\UNTIL{convergence} \label{alg:SCA5}
\STATE \textbf{Output:} $\{\mathbf{b}^*_q\}, \bm{\phi}^*$
\end{algorithmic}
\end{algorithm}

\subsection{Initialization procedure}\label{subsec:initializationProcedure}
A well-known limitation of SCA-based optimization is that the initial points condition the final result \cite{Razaviyayn.2014}. In our case, the interdependency of the phase shifter configurations and the digital beamforming at the feeder may cause the initialization to fail multiple times before finding a feasible solution. This subsection introduces an initialization procedure that guarantees consistent results when implementing Algorithm~\ref{alg:SCA}. 

For a given phase shift configuration $\bm{\phi}$, one can find a feasible set of precoders by minimizing the maximum output power of the RF chains while satisfying the devices' RF power requirements. We formulate this problem in epigraph form as
\begin{subequations}\label{eq:epigraphInit}
    \begin{alignat}{2}
    \mathbf{P3:} \ &\underset{\{\mathbf{b}_q\},t}{\mathrm{minimize}} && \ t\label{eq:P3a} \\
    & \text{s.t.} && \ g\sum_{q=1}^{Q} |b_{q,n}|^2 \leq t, \ \forall n \in \mathcal{N}, \label{eq:P3b} \\
    & \quad && \ P_k \geq P^\mathrm{th}_k, \forall k \in \mathcal{K}, \label{eq:P3c}\\
    & \quad && \ t \geq 0. \label{eq:P3d}
    \end{alignat}
\end{subequations}
We now transform \eqref{eq:P3c} into a convex set of constraint as follows
\begin{align*}
    g \sum_{q=1}^{Q} |\hat{\mathbf{h}}^H_k \mathbf{b}_q|^2 = g \hat{\mathbf{h}}^H_k \sum_{q=1}^{Q} \mathbf{b}_q \mathbf{b}^H_q \hat{\mathbf{h}}_k
    = g \mathrm{tr}(\hat{\mathbf{H}}_k \mathbf{B}), 
\end{align*}
where $\hat{\mathbf{H}}_k = \hat{\mathbf{h}}_k \hat{\mathbf{h}}^H_k$ and $\mathbf{B} = \sum_{q=1}^{Q} \mathbf{b}_q \mathbf{b}^H_q$. Similarly, we transform \eqref{eq:P3b} into
\begin{align*}
    g \sum_{q=1}^Q |b_{q,n}|^2 &= g \mathbf{e}^H_n \sum_{q=1}^Q \mathbf{b}_q \mathbf{b}^H_q \mathbf{e}_n = g \mathrm{tr}(\mathbf{E}_n \mathbf{B}),
\end{align*}
where $\mathbf{E_n} = \mathbf{e}_n \mathbf{e}^H_n$ and $\mathbf{e}_n \in \mathbb{R}^{N \times 1}$ is a standard basis vector with the $n^\mathrm{th}$ entry being nonzero. This mathematical construct ensures the selection of only those entries of the precoders that contribute to the output power of the $n^\mathrm{th}$ RF chain. Then, we reformulate $\mathbf{P3}$ as a semidefinite program (SDP)
\begin{subequations}\label{P4}
    \begin{alignat}{2}
    \mathbf{P4:} \ &\underset{\mathbf{B},t}{\mathrm{minimize}} \ && t \label{P4a}\\
    &\text{s.t.} \ && g\mathrm{tr}(\mathbf{E}_n \mathbf{B}) \leq t, \ \forall n \in \mathcal{N}, \label{P4b}\\
    & \ && g\mathrm{tr}(\hat{\mathbf{H}}_k \mathbf{B}) \geq P^\mathrm{th}_k, \ \forall k \in \cal K, \label{P4c}\\
    & \ && \mathbf{B} \succeq 0, \label{P4d}\\
    & \ && t \geq 0. \label{P4e}
    \end{alignat}
\end{subequations}
Since no relaxations or approximations were introduced in the formulation of $\mathbf{P4}$, it is equivalent to $\mathbf{P3}$. The precoders $\{\mathbf{\Tilde{b}}_q\}$ are then obtained via the eigenvalue decomposition of $\mathbf{B}$, which depending on the channel conditions may yield $Q \leq N$ precoders. Observe that the obtained solution is a feasible initializer of $\mathbf{P2}$ only if $g \sum_{q=1}^Q |b_{q,n}|^2 \leq P_\mathrm{max}, \ \forall n \in \mathcal{N}$.

The initialization of the phase shift configuration problem remains an issue. In fact, according to some experiments, a random initialization strategy for this setup does not provide consistent results that help us identify a clear trend, especially when the number of ITS phase shifters increases. In the sequel, we provide a heuristic algorithm to tackle this limitation. We divide the ITS into $N$ nonoverlapping clusters and assign each cluster to a different feeder's antenna. Let $\mathcal{M}_n \in \mathcal{M}$ represent the set of passive elements associated with to the $n^\mathrm{th}$ subset, defined as $\mathcal{M}_n = \Big\{m \ | \ n = \underset{n'}{\arg\min} \ |a_{n',m}|^2 \Big\}$, thus forming $N$ virtual single-RF chain transmitters. This approach is justified by the observation that each element within the ITS experiences varying gains relative to each feeder's antenna. We note that this approach facilitates initializing the ITS configuration since, for the $m^\mathrm{th}$ element, the phase shift can be computed as $\angle \phi_m = \angle h_{k,m} - \angle a_{n,m}, \ \forall m \in \mathcal{M}_n$, compensating the phase difference between the corresponding active antenna and device channels, and therefore mimicking a maximum ratio transmission. 

Observe that when $N > K$, the number of assigned clusters per device might be different. This variation depends on the influence each virtual single-RF transmitter has on the received power at each device. Let $N_k \geq 1$ denote the number of ITS clusters assigned to the $k^\mathrm{th}$ device. We want that
\begin{equation}
    \varrho_1|\mathbf{h}_1|^2 \approx \ldots \approx \varrho_k|\mathbf{h}_k|^2 \approx \ldots \approx \varrho_K|\mathbf{h}_K|^2,  
\end{equation}
where 
\begin{equation}\label{eq:InitSCAMetric}
    \varrho_k = N^2_k + \sum_{k' \in \mathcal{K} \setminus k}^{K} N_{k'}.
\end{equation}
That is, we aim to find a cluster assignment that yields a similar product $\rho_k|\mathbf{h}_k|^2$ across all devices. Intuitively, this approach considers the effect of the entire ITS while giving greater importance to the impact of each device's assigned cluster compared to clusters assigned to others. It is worth mentioning that this heuristic has been obtained after observing the results of extensive simulations.

Assigning the ITS clusters to the devices is a combinatorial problem since the specific set of clusters assigned per device impacts the final results. Let $\Pi$ denote the set containing all permutations of an $N-$element multiset in which the $k^\mathrm{th}$ device index appears $N_k$ times. Hence, the number of permutations in $\Pi$ is $\frac{N!}{N_1! N_2! \cdots N_K!}$. Each permutation $\pi \in \Pi$ represents the assignment of an ITS cluster to a device, \textit{e.g.}, the $n^\mathrm{th}$ entry of $\pi$ contains the index of the device to which this cluster is assigned. 

The initialization procedure outlined in this subsection is summarized in Algorithm~\ref{alg:InitSCA}. Lines \ref{alg:InitSCA2} to \ref{alg:InitSCA7} allocate a set of ITS clusters to each device, while lines \ref{alg:InitSCA9}-\ref{alg:InitSCA17} evaluate the power consumption for each permutation $\pi \in \Pi$, $\{\Tilde{\mathbf{b}}_q\}$ when solving $\mathbf{P4}$. The permutation that minimizes power consumption is used as the initial guess for running Algorithm~\ref{alg:InitSCA}. Notably, since each permutation's evaluation is independent, this process can be parallelized to reduce processing time. Overall, the complexity of Algorithm~\ref{alg:InitSCA} is dominated by the solution of the SDP in the step \ref{alg:InitSCA11}, which involves $\frac{N(N+1)}{2} + 1$ decision variables and $N + M + 1$ constraints. IPM can find an $\epsilon-$optimal solution in $\mathcal{O}(\sqrt{N})$ iterations each demanding $\mathcal{O}(N^6)$ arithmetic operations \cite[Section 6.6.3]{Ben.2001}. Additionally, for cases where $N < K$ or $N \gg K$, a different initialization method, such as those based on nature-inspired algorithms, may be preferable, but this demands further investigation.
\begin{algorithm}[t!]
\caption{SCA initialization}
\begin{algorithmic}[1] \label{alg:InitSCA}
 \STATE \textbf{Input:} $\{P^\mathrm{th}_k, \mathbf{h}_k\}_{\forall k\in\mathcal{K}}, \mathbf{A}, N, M$ \label{alg:InitSCA1}
 \STATE Set $N_k = 1, \forall k \in \mathcal{K}$ \label{alg:InitSCA2}
\WHILE{$\sum_{k=1}^{K} N_k < N$} \label{alg:InitSCA3}
\STATE Compute $\varrho_k$ using \eqref{eq:InitSCAMetric} \label{alg:InitSCA4}
\STATE $k^* \gets \underset{k}{\mathrm{min}} \ \varrho_k |\mathbf{h}_k|^2$ \label{alg:InitSCA5}
\STATE $N_{k^*} \gets N_{k^*} + 1$ \label{alg:InitSCA6}
\ENDWHILE \label{alg:InitSCA7}
\STATE Set $P = +\infty$ \label{alg:InitSCA8}
\FORALL{$\bm{\pi} \in \Pi$} \label{alg:InitSCA9}
    \STATE Compute $\bm{\phi}$ \label{alg:InitSCA10}
    \STATE Solve $\mathbf{P4}$, output $\{\mathbf{b}_q\}$ \label{alg:InitSCA11}
    \IF {$P > \sum_{n = 1}^N P_{\mathrm{hpa},n}$} \label{alg:InitSCA12}
        \STATE $P \gets \sum_{n = 1}^N P_{\mathrm{hpa},n}$ \label{alg:InitSCA13}
        \STATE $\{\mathbf{\Tilde{b}}_q\} \gets \{\mathbf{b}_q\}$ \label{alg:InitSCA14}
        \STATE $\bm{\Tilde{\phi}} \gets \bm{\phi}$ \label{alg:InitSCA15}
    \ENDIF \label{alg:InitSCA16}
\ENDFOR \label{alg:InitSCA17}
    \STATE \textbf{Output:} $\{\mathbf{\Tilde{b}}_q\}, \bm{\Tilde{\phi}}$ \label{alg:InitSCA18}
\end{algorithmic}
\end{algorithm}

\section{Benchmark technologies}\label{sec:benchmarkSolutions}
\begin{figure}
    \centering
    \includegraphics[width=\linewidth]
    {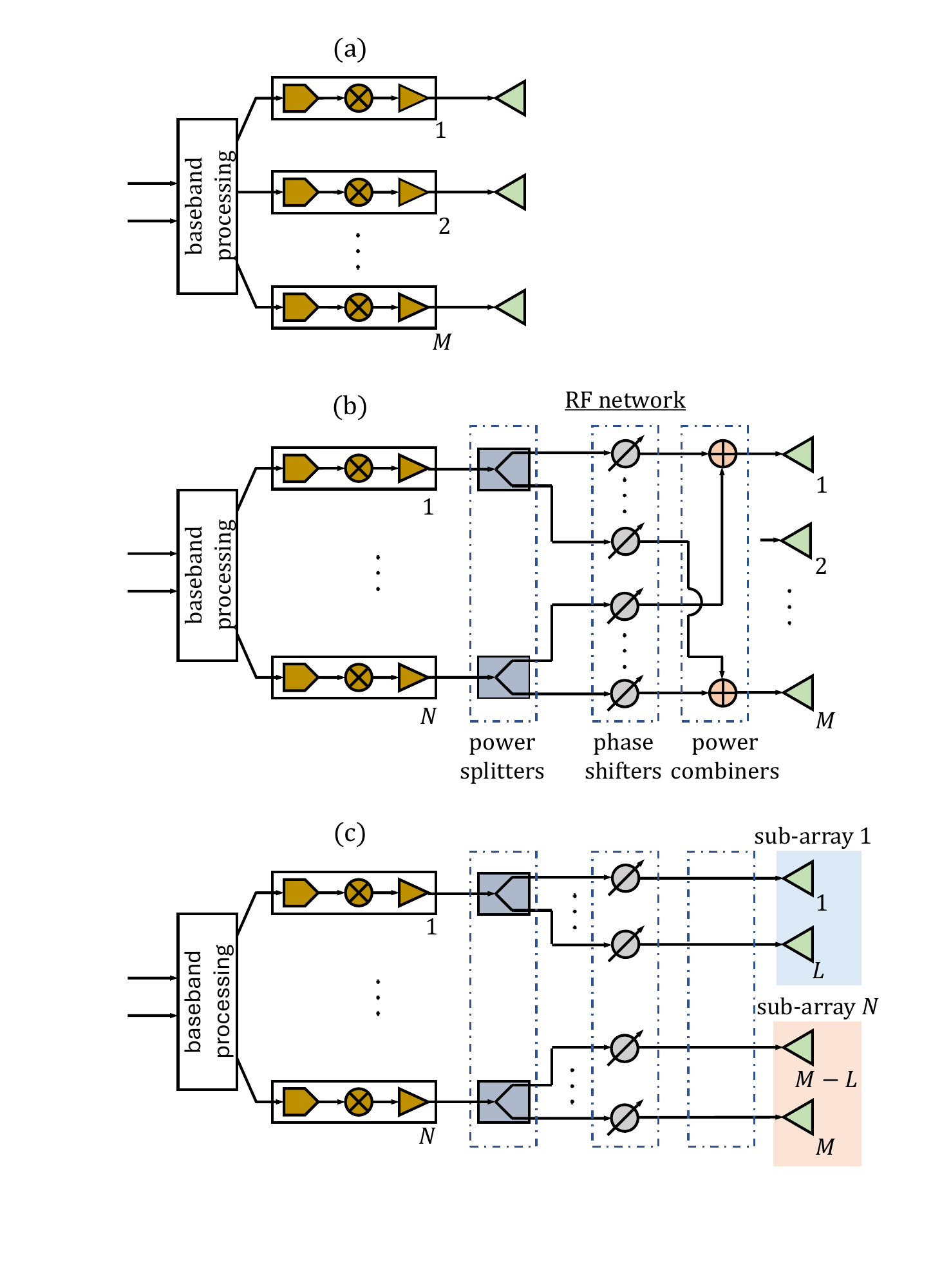}
    \vspace{-1.5em}
    \caption{Benchmark beamforming architectures implemented on the PB, including: (a) FD, (b) HBFC, and (c) HBPC.}
    \label{fig:benchmarkArchitectures}
\end{figure}

In this section, we develop three benchmark solutions based on fully-digital (FD) beamforming and hybrid analog-digital beamforming architectures, including both hybrid partially-connected (HBPC) and hybrid fully-connected (HBFC) designs, without equipping an ITS. Therefore, we set the number of transmit antennas equal to the number of passive elements in the ITS-assisted architecture. As for the ITS-equipped PB, herein we assume that the $P_\mathrm{bb}$ and $P_\mathrm{tc}$ are independent of the optimization variables. The implementation of the benchmark architectures is illustrated in Fig.~\ref{fig:benchmarkArchitectures}, with further details provided in the subsequent subsections.
\vspace{-0.7em}
\subsection{FD beamforming architecture}\label{subsec:digitalBeamforming}
We assume that the transmitter architecture equipped with FD beamforming, which is shown in \ref{fig:benchmarkArchitectures}a, consists of $M$ antennas connected to dedicated RF chains. For simplicity, we will use the same notation as for the digital processing section of the ITS-assisted transmitter. Then, the received signal at the $k^\mathrm{th}$ device is  $y_k = \sqrt{g}\mathbf{h}^H_k \sum_{q=1}^Q \mathbf{b}_q s_q$, where $\mathbf{b}_q \in \mathbb{C}^{M \times 1}$ and $Q \leq M$. Hence, the power received at the $k^\mathrm{th}$ device is $P_k = g \sum_{q=1}^Q |\mathbf{h}^H_k \mathbf{b}_q|^2$. We model the power consumption of the digital beamforming architecture as
\begin{equation}
    P^\mathrm{fd}_T = P_\mathrm{bb} + P_\mathrm{tc} + \sum_{m=1}^M P_{\mathrm{hpa},m},
\end{equation}
where $P_{\mathrm{hpa},m}$ is determined by \eqref{eq:dohertyPowConsumpt} for an input power computed using \eqref{eq:inputPowDoherty}. Hence, we can formulate the following optimization problem to minimize the total power consumption of the transmitter as
\begin{subequations}\label{P5}
    \begin{alignat}{2}
    \mathbf{P5:} \ &\underset{\{\mathbf{b}_q\}}{\mathrm{minimize}} \ && \sum_{m=1}^M P_{\mathrm{hpa},m} \label{P5a}\\
    &\text{s.t.} \ && g \sum_{q=1}^{Q} |\mathbf{h}^H_k \mathbf{b}_q|^2 \geq P_k^\mathrm{th}, \ \forall k \in \cal K. \label{P5b}
    \end{alignat}
\end{subequations}
Following the procedure in Section~\ref{sec:proposedSolution}, we can find a suboptimal solution of $\mathbf{P5}$ by solving a series of SOCPs of the form
\begin{subequations}\label{P6}
    \begin{alignat}{2}
    \mathbf{P6:} \ &\underset{\{\mathbf{b}_q\},\{t_m\}}{\mathrm{minimize}} \ && \sum_{m = 1}^{M} t_m \label{P6a}\\
    &\text{s.t.} \ && \eqref{eq:branch1.1},\eqref{eq:branch1.2}, \ \{\forall m \in \mathcal{M} \ | \ \eqref{eq:cond1} \}, \label{P6b}\\
    & \ && \eqref{eq:branch2.1},\eqref{eq:branch2.2},\eqref{eq:branch2.3approx}, \ \{\forall m \in \mathcal{M} \ | \ \eqref{eq:cond2} \}, \label{P6c} \\
    & \ && g\sum_{q = 1}^{Q} \Tilde{P}_{k,q} \geq P_k^\mathrm{th} , \ \forall k \in \cal K, \label{P6d}
    \end{alignat}
\end{subequations}
where either constraint \eqref{P6b} or \eqref{P6b} is activated depending on the value of $\Tilde{b}_{q,m}$ and \eqref{eq:cond1} and \eqref{eq:cond2} are evaluated with the subindex $m$. Meanwhile, we take $P_{k,q} = 2\Re\left\{z^H_{k,q}\mathbf{h}^H_k \mathbf{b}_q\right\} - |z_{k,q}|^2$ in \eqref{P6d}. Moreover, the initial points $\{\Tilde{b}_{q,m}\}$ can be obtained from solving $\mathbf{P4}$ for $N = M$. Consequently, initialization procedure using IPM requires $\mathcal{O}(\sqrt{M})$ iterations and $\mathcal{O}(M^6)$ arithmetic operations when solving $\mathbf{P4}$ \cite[Section 6.6.3]{Ben.2001}. Moreover, solving $\mathbf{P6}$ with $3M + K$ constraints and $2M$ variables necessitates $\mathcal{O}(\sqrt{M})$ iterations each costing $\mathcal{O}(M^{3.5})$ arithmetic operations when using IPM \cite[Section 6.6.2]{Ben.2001}.

\subsection{Hybrid analog-digital beamforming architectures}
\label{subsec:hybridBeamforming}
We assume that both the HBFC and HBPC beamforming architectures are composed of $N$ RF chains and $M$ transmit antennas as depicted in Fig.~\ref{fig:benchmarkArchitectures}b and Fig.~\ref{fig:benchmarkArchitectures}c, respectively. Unlike the ITS-equipped architecture, the hybrid analog-digital beamforming transmitters incorporate an RF network---comprising power splitters, phase shifters, and power combiners---connecting the output of the HPAs to the corresponding antennas. In the HBFC architecture, the output of each RF chain is distributed to all antennas in the array, whereas in the HBPC architecture, each RF chain feeds a distinct, nonoverlapping subset of antennas. This distinction influences not only the number of phase shifters required for each architecture---$M$ and $MN$ in the HBPC and HBFC architectures, respectively---but also the number of outputs (inputs) at the power dividers (combiners). Observe that the architectures herein presented provide a low hardware complexity since the number of HPAs do not scale with the number of antennas \cite{Karacora.2019}. The aforementioned factors, in turn, impact the insertion losses of the RF network, as we will explain next.

Let $\gamma$ denote the total insertion losses introduced by the RF network to the signal path. Then, the received signal at the $k^\mathrm{th}$ device is $y_k = \sqrt{\frac{g}{\gamma}}\mathbf{h}^H_k \mathbf{C} \sum_{q=1}^Q \mathbf{b}_q s_q$, where $\mathbf{C}$ is the analog precoder. Consequently, the RF power recieved at the $k^\mathbf{th}$ device is $P_k = \frac{g}{\gamma} \sum_{q=1}^Q |\mathbf{h}^H_k \mathbf{C} \mathbf{b}_q|^2$. For the case of HBFC, $\mathbf{C}$ is a full matrix, \textit{i.e.}, $\mathbf{C} = [\mathbf{c}_1, \ \mathbf{c}_2, \ \ldots, \ \mathbf{c}_n, \ \ldots, \ \mathbf{c}_N]$, where $\mathbf{c}_n \in \mathbb{C}^{M\times1}$, whose $m^\mathrm{th}$ entry is $c_{n,m} = \frac{1}{\sqrt{M}}e^{j\alpha_{n,m}}$. Meanwhile, for the case of HBPC,  $\mathbf{C} = \mathrm{blkdiag}(\mathbf{c}_1, \ \mathbf{c}_2, \ \ldots, \ \mathbf{c}_n, \ \ldots, \ \mathbf{c}_N)$ is a block diagonal matrix where $\mathbf{c}_n \in \mathbb{C}^{L\times1}$, whose $l-$entry is $c_{n,l} = \frac{1}{\sqrt{L}}e^{j\alpha_{n,l}}$, and $L = \lfloor \frac{M}{N} \rfloor$ is the number of antennas in the sub-array connected to the $n^\mathrm{th}$ RF chain. Therefore, the total insertion losses for the HBFC and the HBPC architectures are $\lceil \log_2{M} \rceil\gamma_s \lceil \log_2{N} \rceil\gamma_c\gamma_p$ and $\lceil \log_2{L} \rceil\gamma_s\gamma_p$, respectively, where $\gamma_s$, $\gamma_c$, and $\gamma_p$ denote the insertion losses of the power splitters, power combiners, and phase shifters. We model the power consumption of the hybrid architectures as follows
{\setlength{\abovedisplayskip}{4pt}
 \setlength{\belowdisplayskip}{4pt}
\begin{equation}
    P_T^\mathrm{hb} = P_\mathrm{bb} + P_\mathrm{tc} + \sum_{n=1}^N P_{\mathrm{hpa},n},
\end{equation}
}
where $P_{\mathrm{hpa},n}$, determined by \eqref{eq:dohertyPowConsumpt}, differentiates the power consumption of the HBFC and HBPC architectures. The corresponding power minimization problem is formulated as
{\setlength{\abovedisplayskip}{4pt}
 \setlength{\belowdisplayskip}{4pt}
\begin{subequations}\label{P7}
    \begin{alignat}{2}
    \mathbf{P7:} \ &\underset{\{\mathbf{b}_q\}, \mathbf{C}}{\mathrm{minimize}} \ && \sum_{n=1}^N P_{\mathrm{hpa},n} \label{P7a}\\
    &\text{s.t.} \ && g \sum_{q=1}^{Q} |\mathbf{h}^H_k \mathbf{C} \mathbf{b}_q|^2 \geq P_k^\mathrm{th}, \ \forall k \in \cal K, \label{P7b} \\
    & \ && |c_{n,m}| = \Delta, \forall n \in \mathcal{N}, \forall m \in \mathcal{M}, \label{P7c}
    \end{alignat}
\end{subequations}
}
where $\Delta \in \left\{\frac{1}{\sqrt{L}}, \frac{1}{\sqrt{M}} \right\}$ depends on the particular hybrid beamforming implementation. To deal with the nonconvex received power requirement, we can take $\mathbf{\hat{h}}^H_k = \mathbf{h}^H_k \mathbf{C}$ and apply the same procedure as in \eqref{eq:rPowApprox} to obtain a lower bound of the received power constraint. Meanwhile, the objective function can be transformed into a set of convex constraints following the procedure described in Section~\ref{sec:proposedSolution}. This allows us to find a suboptimal solution of $\mathbf{P7}$ by solving a series of problems in the form
\begin{subequations}\label{P8}
    \begin{alignat}{2}
    \mathbf{P8:} \ &\underset{\{\mathbf{b}_q\},\mathbf{C},\{t_n\}}{\mathrm{minimize}} \ && \sum_{n=1}^{N} t_n \label{P8a}\\
    &\text{s.t.} \ && \eqref{eq:branch1.1},\eqref{eq:branch1.2}, \ \{\forall n \in \mathcal{N} \ | \ \eqref{eq:cond1} \}, \label{P8b}\\
    & \ && \eqref{eq:branch2.1},\eqref{eq:branch2.2},\eqref{eq:branch2.3approx}, \ \{\forall n \in \mathcal{N} \ | \ \eqref{eq:cond2} \}, \label{P8c} \\
    & \ && \frac{g}{\gamma}\sum_{q = 1}^{Q} \Tilde{P}_{k,q} \geq P_k^\mathrm{th} , \ \forall k \in \cal K, \label{P8d}\\
    & \ && |c_{n,m}| \leq \Delta, \forall n \in \mathcal{N}, \forall m \in \mathcal{M}, \label{P8e}\\
    & \ && t_n \geq 0, \ \forall n \in \mathcal{N}, \label{P8f}
    \end{alignat}
\end{subequations}
where either constraint \eqref{P8b} or \eqref{P8c} is evaluated depending on the value of $\Tilde{b}_{q,n}$. Notice that in \eqref{P8e} we have relaxed the constant modulus constraint for the entries of $\mathbf{C}$, which will hold with equality at the optimal point. Now, to obtain the initial $\{\mathbf{\Tilde{b}_q}\}$ and $\mathbf{\Tilde{C}}$ to solve problem $\mathbf{P8}$, we propose Algorithm~\ref{alg:InitSCAHB}. It begins by determining the number of RF chains assigned to each device, which, in this case, corresponds to assigning $N_k$ columns in matrix $\mathbf{\Tilde{C}}$. Then, the phase of the entries in the columns assigned to the $k^\mathrm{th}$ device are computed as $c_{n,l} = \angle h_{k,l}, \ l = 1, 2, \ldots, L$ for HBPC, and $c_{n,m} = \angle h_{k,m}, \ m = 1, 2, \ldots, M$ for for HBFC. It is worth mentioning that the proposed initialization algorithm has the same worst-case complexity as Algorithm~\ref{alg:InitSCA}, since its most demanding step involves solving $\mathbf{P4}$, which for this case has the same dimensions that for the ITS-equipped PB case. 

Overall, the most critical step in the proposed solution involves solving a SOCP reformulation of $\mathbf{P8}$. For the HBFC architecture, the problem involves $N(4 + M) + K$ constraints and $N^2 + N(M+1)$ variables, whereas for the HBPC case, it involves $N(L + 4) + K$ constraints and $N^2 + N(L+1)$ variables. Given that $M \gg N$, the worst-case complexity of solving this problem using IPM requires $\mathcal{O}(\sqrt{MN})$ and $\mathcal{O}(\sqrt{LN})$ iterations for the HBFC and the HBPC architectures, respectively. Each iteration involves $\mathcal{O}(N^{3.5}M^{3.5})$ and $\mathcal{O}(N^{3.5}L^{3.5})$ arithmetic operations for the HBFC  and the HBPC, respectively, as detailed in \cite[Section 6.6.2]{Ben.2001}.

\begin{algorithm}[t!]
\caption{SCA initialization for the hybrid beamforming architectures}
\begin{algorithmic}[1] \label{alg:InitSCAHB}
 \STATE \textbf{Input:} $\{P^\mathrm{th}_k, \mathbf{h}_k\}_{\forall k\in\mathcal{K}}, N, M$ \label{alg:InitSCAHB1}
 \STATE Set $N_k = 1, \forall k \in \mathcal{K}$ \label{alg:InitSCAHB2}
\WHILE{$\sum_{k=1}^{K} N_k < N$} \label{alg:InitSCAHB3}
\STATE Compute $\varrho_k$ using \eqref{eq:InitSCAMetric} \label{alg:InitSCAHB4}
\STATE $k^* \gets \underset{k}{\mathrm{min}} \ \varrho_k |\mathbf{h}_k|^2$ \label{alg:InitSCAHB5}
\STATE $N_{k^*} \gets N_{k^*} + 1$ \label{alg:InitSCAHB6}
\ENDWHILE \label{alg:InitSCAHB7}
\STATE Compute $\mathbf{\Tilde{C}}$ \label{alg:InitSCAHB8}
\STATE Solve $\mathbf{P4}$ using $\mathbf{\Tilde{h}}^H_k = \mathbf{h}^H_k \mathbf{C}$, output $\{\mathbf{\Tilde{b}}_q\}$ \label{alg:InitSCAHB9}
\STATE \textbf{Output:} $\{\mathbf{\Tilde{b}}_q\}, \bm{\Tilde{\phi}}$ \label{alg:InitSCAHB10}
\end{algorithmic}
\end{algorithm}

\begin{table*}[t!]
    \caption{Comparison among different beamforming implementations.}
    \centering
    \begin{tabular}{c|c|c|c|c|c}
        \thickhline
        \textbf{Architecture} & \textbf{Precoders} & \textbf{HPAs} & \begin{tabular}{c} 
            \textbf{Power dividers/phase} \\
            \textbf{shifters/power combiners}
            \end{tabular} & \textbf{PB's losses components} & \begin{tabular}{c}
            \textbf{Solution complexity} \\
            \textbf{(iterations, operations)}
            \end{tabular} \\
        \hline
        FD & $\{\mathbf{b}_q\}$ & $M$ & - & $\eta_\mathrm{max}$ & $\mathcal{O}(\sqrt{M})$, $\mathcal{O}(M^{3.5})$ \\
        HBFC & $\{\mathbf{b}_q\}$, $\mathbf{C}$ & $N$ & $N/MN/N$ & $\lceil \log_2{M} \rceil\gamma_s \lceil \log_2{N} \rceil\gamma_c\gamma_p$, $\eta_\mathrm{max}$ & $\mathcal{O}(\sqrt{MN})$,  $\mathcal{O}(N^{3.5}M^{3.5})$  \\
        HBPC & $\{\mathbf{b}_q\}$, $\mathbf{C}$ & $N$ & $N/M/-$ & $\lceil \log_2{L} \rceil\gamma_s\gamma_p$, $\eta_\mathrm{max}$ & $\mathcal{O}(\sqrt{LN})$, $\mathcal{O}(N^{3.5}L^{3.5})$ \\
        ITS-equipped & $\{\mathbf{b}_q\}$, $\bm{\phi}$ & $N$ & $-/M/-$ & $\mathbf{A}$, $\rho_\mathrm{its}$, $\eta_\mathrm{max}$ & $\mathcal{O}(\sqrt{N})$, $\mathcal{O}(N^6)$ \\
        \thickhline
    \end{tabular}
    \label{tab:comparisonBeamformingArchitectures}
\end{table*}

\section{Numerical Analysis}\label{sec:numericalAnalysis}
\begin{figure}[t!]
    \centering
    \includegraphics[width=\linewidth]{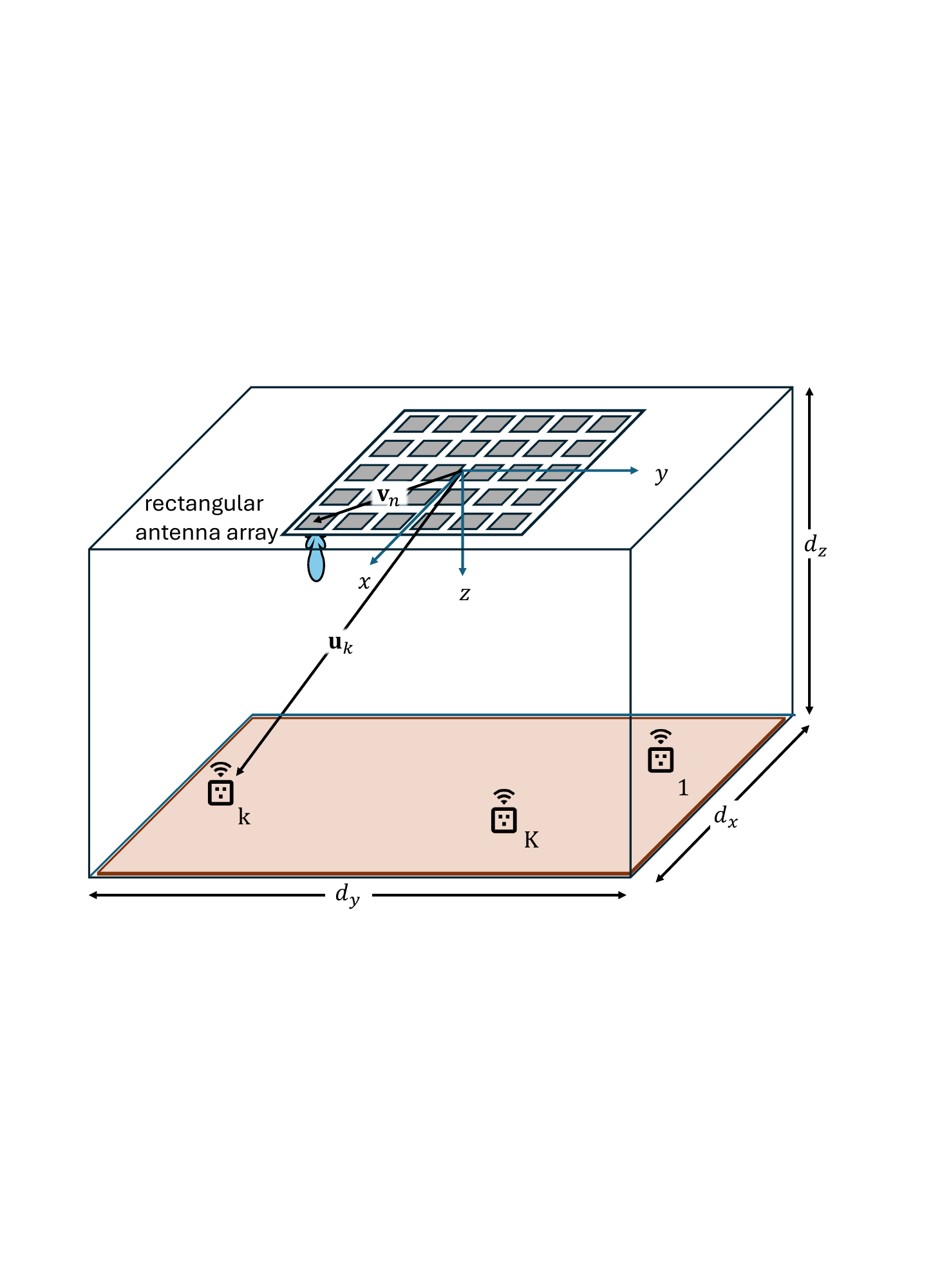}
    \vspace{-1em}
    \caption{Simplified representation of the simulation setup. The rectangular array on top represents the ITS and the conventional antenna array for the benchmark technologies.}
    \label{fig:simulationSetup}
\end{figure}

\begin{figure}[t!]
    \centering
    \includegraphics[width=\linewidth]{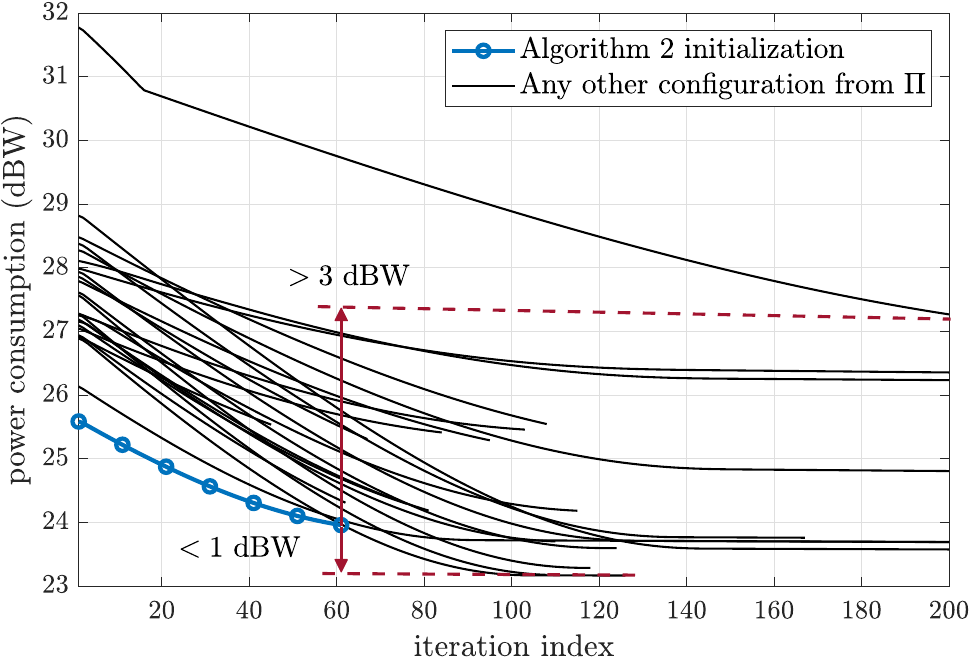}
    \vspace{-1.5em}
    \caption{Convergence rate of Algorithm~\ref{alg:SCA} for different initializations of $\bm{\phi}$ and $\{\mathbf{b}_q\}$ using all the RF chains to devices assignments from the set $\Pi$. We note that when $N = 4$ and $M = 100$, $\Pi$ contains $24$ permutations.}
    \label{fig:convergence}
\end{figure}

In this section, we analyze the performance of the proposed ITS-equipped PB and the discussed benchmark technologies. First, Table~\ref{tab:comparisonBeamformingArchitectures} presents a comparison of the distinctive features of the aforementioned beamforming architectures based on various criteria. It can be observed that the proposed PB architecture scales favorably with the number of antennas in terms of the required number of HPAs and associated losses. Furthermore, the computational complexity of determining the configuration for the ITS-equipped PB requires the fewest iterations assuming $M \gg N$, although the per-iteration cost exhibits the least favorable scaling behavior. Then, Fig.~\ref{fig:simulationSetup} illustrates the simulation setup where the devices are uniformly distributed in a plane with dimensions $d_x \times d_y$ located at a distance $d_z$ from the ITS. By default, we set the dimensions of the service area to $d_x = d_y = 3~$m and $d_z = 5~$m, with its center axially aligned with that of the ITS along the $z-$axis. All results are obtained by averaging over $100$ deployment realizations. Unless otherwise stated, we utilize the simulation parameters listed in Table~\ref{tab:param}. Values of $r_a$ and $\delta_a$, computed in accordance with \cite{Jamali.2021}, are chosen to ensure adequate separation between radiating elements and hence minimize mutual coupling.
\begin{table}[t!]
    \centering
    \caption{Simulation Parameters \cite{Pei.2021, Jamali.2021,Rosabal.2024,Garcia.2016}}
    \label{tab:param}
    \begin{tabular}{c c | c c | c c}
        \thickhline
            \textbf{Parameter} & \textbf{Value} & \textbf{Parameter} & \textbf{Value} \\
        \hline
            $P_\mathrm{bb}$ & $200~$mW & $K$ & $4$ \\
            $P_\mathrm{tc}$ & $100~$mW & $M$ & $100$ \\
            $P_\mathrm{ctrl}$ & $1~$W & $N$ & $4$ \\
            $P_\mathrm{cell}$ & $1~$mW & $f_c$ & $5~$GHz \\
            $P^\mathrm{th}_k$ & $1~$mW & $\gamma_s$ & $0.5~$dB \\
            $P_\mathrm{max}$ & $300~$W & $\gamma_c$ & $0.5~$dB \\
            $\mu$ & $10$ & $\gamma_p$ & $3.5~$dB \\
            $\kappa$ & $2$ & $r_a$ & $\frac{\lambda}{2 \sin{(\pi/N)}}$ \\
            $\delta$ & $\frac{\lambda}{2}$ & $\delta_a$ & \textcolor{blue}{$2\lambda\sqrt{\frac{M}{\pi}}$} \\
            $\eta_\mathrm{max}$ & $0.25$ & $\ell$ & $2$ \\
            $\rho_\mathrm{its}$ & $0.45$ & $g$ & $100$ \\
        \thickhline
    \end{tabular}
\end{table} 
Fig.~\ref{fig:convergence} shows how the convergence rate of Algorithm~\ref{alg:SCA} changes for  different initializations of $\bm{\phi}$ and $\{\mathbf{b}_q\}$, as highlighted in Section~\ref{subsec:initializationProcedure}. Observe that Algorithm~\ref{alg:SCA} achieves a favorable trade-off between convergence speed and solution quality when utilizing Algorithm~\ref{alg:InitSCA}. In fact, for this scenario, the gap between the final power consumption obtained with the proposed initialization and that of the best-performing initializations is relatively small, especially when compared to the larger gap observed with poorly performing initial points. 

\begin{figure}[t!]
    \centering
    \includegraphics[width=\linewidth]{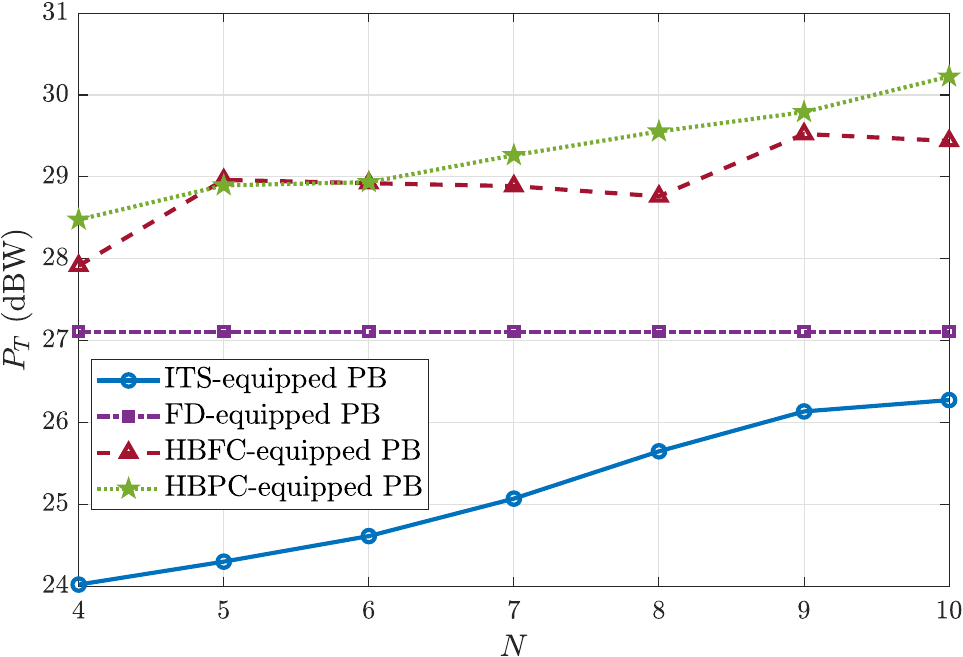}
    \vspace{-1.5em}
    \caption{Power consumption as a function of the number of RF chains. For the FD-equipped PB, we consider the number of RF chains to be $M = 100$.}
    \label{fig:powVSN}
\end{figure}

Fig.~\ref{fig:powVSN} illustrates the power consumption of the PB as a function of the number of RF chains. Results evince that the ITS-equipped PB outperforms the benchmark technologies; however, its power consumption approaches that of the FD-equipped PB where the number of RF chains is $M = 100$. Notably, in this configuration, the FD-equipped PB achieves the best performance among all benchmark technologies. Moreover, power consumption increases with the number of RF chains for all technologies---except for FD, whose results are independent of $N$. This trend confirms that adding more RF chains generally increases the PB's power demands as more HPAs are required. Lastly, the results for the hybrid architectures do not follow a steady increasing pattern. This behavior arises from the losses model described in Section~\ref{subsec:hybridBeamforming}, where, due to the ceiling operation, losses remain unchanged over certain RF chain ranges (\textit{e.g.}, from $5$ to $8$ for the HBFC), leading to reduced power consumption.

\begin{figure}[t!]
    \centering
    \includegraphics[width=\linewidth]{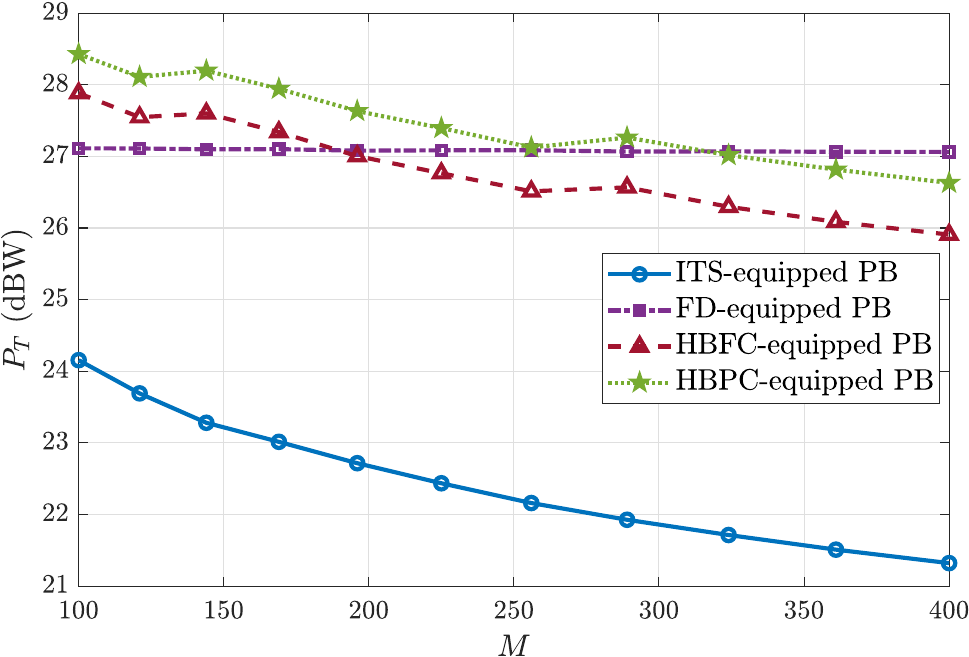}
    \vspace{-1.5em}
    \caption{Power consumption as a function of the number of antennas/ITS elements. Notice that the effective number of transmit antennas for HBPC can be less than $M$, \textit{i.e.}, $NL < M$. Additionally, we set $N = 4$ for all configurations except for the FD case, where $M$ RF chains are employed and the corresponding power consumption is $27~$dBW.}
    \label{fig:powVSM}
\end{figure}

Fig.~\ref{fig:powVSM} illustrates the relationship between power consumption and the number of transmit antennas, showing a general decline in consumption as the number of antennas increases. This trend is attributed to the enhanced array gain, which enables lower transmit power while maintaining the desired performance. Notably, the FD-equipped PB exhibits the smallest reduction due to its large number of RF chains. Additionally, the performance gap between the ITS-equipped PB, which is the best-performing architecture, widens with the number of antennas, reinforcing its effectiveness in handling large number of antennas. Furthermore, for $M < 196$, the FD architecture exhibits the lowest power consumption among the benchmark technologies. However, as the number of antennas increases, this trend reverses, and the hybrid architectures achieve lower power consumption for $M \geq 196$ and $M \geq 324$ for HBFC and HBPC, respectively. It is also wort noting that the exact number of transmit antennas in HBPC is lower than the value indicated on the $x-$axis of Fig.~\ref{fig:powVSM}. This discrepancy arises from the rounding operation described in Section~\ref{sec:benchmarkSolutions}, where the number of antennas assigned to each RF chain is determined as $L = \lfloor \frac{M}{N} \rfloor$ for the HBPC architecture. Finally, in this setup, the array size varies from $27~$cm (for $M = 100$) to $57~$cm (for $M = 400$) in side length.

\begin{figure}[t]
    \centering
    \includegraphics[width=\linewidth]{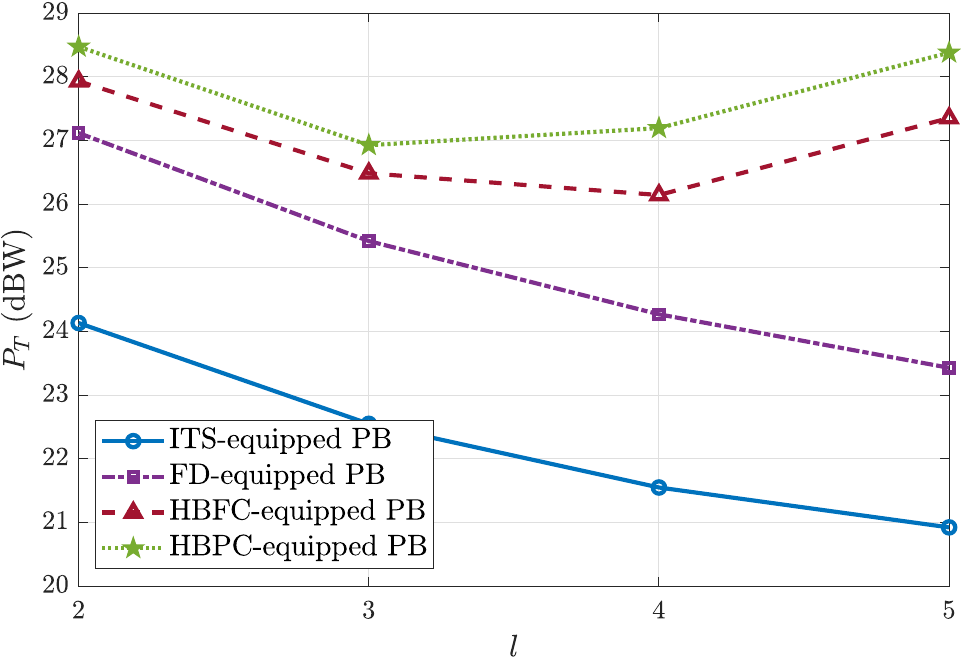}
    \vspace{-1.5em}
    \caption{Power consumption as a function of the number of HPAs in the Doherty architecture for $N=4$ and $M=100$.}
    \label{fig:powVSl}
\end{figure}

Fig.~\ref{fig:powVSl} presents the PB's power consumption against the number of HPAs in the Doherty architecture. Based on Fig.~\ref{fig:DohertyAmplifier}b, one can infer that increasing $\ell$ enhances the efficiency of individual HPAs, leading to a reduction in the overall power consumption. Among the evaluated architectures, the ITS-equipped PB has the lowest power demands, with its consumption decreasing as $\ell$ increases. Interestingly, an inflection point is observed at  $\ell = 4$ and $\ell = 3$ for the HBFC and the HBPC architectures, respectively, highlighting efficiency trade-offs in hybrid beamforming configurations. This is because, as $\ell$ increases, the threshold for activating the peaking HPAs decreases, resulting in all HPAs being active when the transmit power is high. The HBFC and HBPC architectures operate at higher transmit power levels compared to FD and ITS-equipped PBs in this setup. However, this increase is not sufficient to drive all active HPAs into their most efficient operating region. In fact, right after the inflection point, where the peaking HPAs become active, the transmit power of these hybrid architectures slightly decreases, aligning more closely with the transition point where peak efficiency is achieved. 

\begin{figure}[t]
    \centering
    \includegraphics[width=\linewidth]{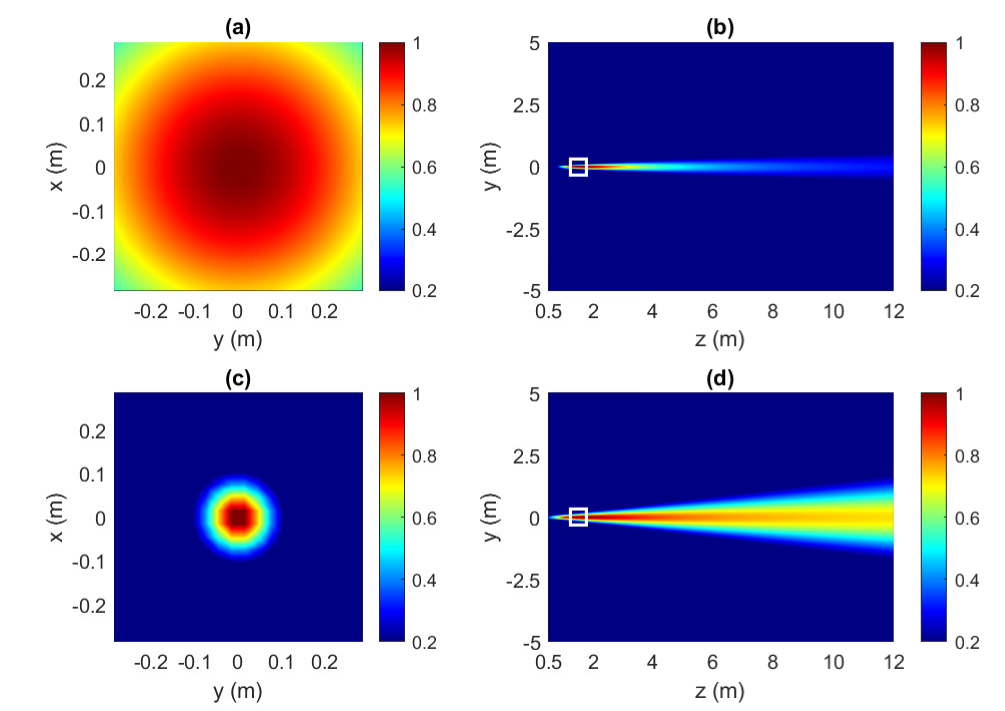}
    \vspace{-1.5em}
    \caption{Normalized received power at the ITS (subfigures a and c) and in the area where the device is located (subfigures b and d). In this experiment, the device is positioned along the $z$-axis at a distance of $1.5~\text{m}$ from the ITS. We consider a single-RF chain PB, \textit{i.e.}, $r_a = 0$, located at $1.35~\text{m}$ (subfigures a and b) and $0.2~\text{m}$ (subfigures c and d) from the ITS. All subfigures are normalized using a global factor corresponding to the maximum received power across all cases. Moreover, the ITS has $400$ elements.}
    \label{fig:nfResultsPart1}
\end{figure}

Finally, Fig.~\ref{fig:nfResultsPart1} demonstrates how the geometry of the ITS-equipped PB influences the propagation conditions. In this scenario, a single-RF chain PB is utilized to charge one device. The full illumination strategy results in the central elements of the ITS receiving most of the impinging power, leading to a nonuniform power distribution across the surface. This effect becomes more pronounced as the feeder gets closer to the ITS, significantly reducing the power received by the edge elements. Consequently, the effective aperture of the PB is modified, causing the device to operate in either the near-field (Fig.~\ref{fig:nfResultsPart1}b) or the far-field (Fig.~\ref{fig:nfResultsPart1}d), depending on the feeder's position, even with the same physical aperture of the ITS. More details about the effects of nonuniform power distribution on antenna apertures can be found in \cite{Xiao.2023}.

\section{Conclusions}\label{sec:conclusions}
This paper studied the power consumption of a PB, equipped with an intelligent surface, charging a network of low-power IoT devices. For this purpose, we characterized the power consumption of the PB considering the nonlinearities of the Doherty HPAs, the power required to control the ITS, and the inherent losses of the air interface between the ITS and the feeder. We utilized SCA to recast the minimization of the power consumption as a series of problems optimizing the feeder's digital precoder and the ITS's phase shifter configuration, simultaneously. Moreover, we provided an initialization algorithm that divides the PB into multiple virtual single-RF chain transmitters utilizing nonoverlapping subsets of ITS elements. Different assignments of virtual transmitters to devices are explored and the one that minimizes power consumption is selected. This strategy achieved a good balance between convergence speed and solution quality, and provided a more reliable methodology to obtain a feasible initial point compared to random initializations. Numerical results evinced the advantages of the proposed ITS-equipped PB against benchmark technologies equipping digital and hybrid analog-digital beamforming architectures. Specifically, we illustrated how the power consumption of the ITS-equipped PB scales with both the number of RF chains and transmit antennas. Furthermore, we showed that, even when the physical aperture of the ITS remains constant, the PB can operate in either the near-field or far-field at the same deployment location, depending on how power is distributed across the ITS.

Future work may extend this study along several avenues. Evaluating alternative illumination strategies, including different arrangements of feeder antennas, could further enhance power efficiency. Incorporating discrete phase shifts into the ITS design would help explore the trade-offs between implementation complexity and achievable performance in practical scenarios. Additionally, optimizing the feeder's physical size and its relative position with respect to the ITS presents an opportunity to improve the energy-focusing capabilities of the PB and explore the impact on the devices' channels. Developing robust charging algorithms that consider imperfect or limited CSI would also broaden the applicability of the proposed system to dynamic real-world environments. Lastly, addressing mutual coupling arising from the close proximity of the feeder to the ITS and sub-wavelength spacing of ITS elements by leveraging multiport network theory \cite{Nerini.2024} constitutes a promising research direction, though it requires a mathematical formulation distinct from that presented here.

\bibliographystyle{IEEEtran}
\bibliography{IEEEabrv,references}
\end{document}